# *IN SILICO* IDENTIFICATION OF CLINICALLY APPROVED MEDICINES AGAINST THE MAIN PROTEASE OF SARS-COV-2, CAUSATIVE AGENT OF COVID-19

**ESTARI MAMIDALA[1]*, RAKESH DAVELLA[2], SWAPNA GURRAPU[3] AND PUJALA SHIVAKRISHNA[4]**

[1-4] *Infectious Diseases Research Lab, Department of Zoology, Kakatiya University, Warangal-506009, Telangana, Indiau, India.*

## ABSTRACT

The COVID-19 pandemic triggered by SARS-CoV-2 is a worldwide health disaster. Main protease is an attractive drug target among coronaviruses, due to its vital role in processing the polyproteins that are translated from the viral RNA. There is presently no exact drug or treatment for this diseases caused by SARS-CoV-2. In the present study, we report the potential inhibitory activity of some FDA approved drugs against SARS-CoV-2 main protease by molecular docking study to investigate their binding affinity in protease active site. Docking studies revealed that drug Oseltamivir (anti-H1N1 drug), Rifampin (anti-TB drug), Maraviroc, Etravirine, Indinavir, Rilpivirine (anti-HIV drugs) and Atovaquone, Quinidine, Halofantrine, Amodiaquine, Tetracylcine, Azithromycin, hydroxycholoroquine (anti-malarial drugs) among others binds in the active site of the protease with similar or higher affinity. However, the *in-silico* abilities of the drug molecules tested in this study, further needs to be validated by carrying out *in vitro* and *in vivo* studies. Moreover, this study spreads the potential use of current drugs to be considered and used to comprise the fast expanding SARS-CoV-2 infection.

## INTRODUCTION

A new occurred human coronavirus (COVID-19) is informed in December 2019 in Wuhan, China.[1,2] Afterward, the COVID-19 underway spreading across the world, putting the whole world on high attentive.[3] The World Health Organization (WHO) surveillance draft in January 2020 stated that, any traveller to Wuhan, Hubei Province in China, two weeks earlier the onset of the signs, is supposed to be a COVID-19 case.[4,5] On April 21st, 2020, a total of 2,31,4621 confirmed infections were reported worldwide, with 1,57,847 deaths with a increasing mortality rate of >4.3%.[5] The World Health Organization (WHO) strategy to comprise the distribution includes the decrease of human-to-human dispersal by preventing the interaction between individuals, accordingly avoiding transmission extension events and interactive critical risk evidence to all populations.[5] In India, the first case of the COVID-19 was reported in Kerala on 30 January 2020. As of 21 April, 2020, there are 14,255 cases and 559 deaths as reported by the Ministry of Health and Family Welfare, Government of India.[6]

COVID-19 is a member of Betacoronaviruses similar to the Severe Acute Respiratory Syndrome Human coronavirus (SARS HCoV) and the also the Middle-East Respiratory Syndrome Human coronavirus (MERS HCoV).[7] Main protease (Mpro) is the one of the greatest-characterized drug targets among coronaviruses.[8] Beside with the papain-like protease(s), this main protease enzyme is vital for processing the polyproteins that are translated from the viral RNA.[8] This crucial function of main protease in virus duplication makes this enzyme a capable target for the expansion of inhibitors and possible treatment remedy for infection of novel coronavirus. To date, no precise therapeutic medicine or vaccine has been accepted for the management of human coronavirus.

Some clinical trials, it has been described that anti-HIV drugs and chloroquine phosphate, an anti-malarial drug, has a assured therapeutic effect on the COVID-19.[9] In specific, chloroquine phosphate is suggested to treat COVID-19 related pneumonia in larger inhabitants in the future. Subsequently, other clinical trials suggested that hydroxychloroquine supplementary with azithromycin is very operative in the management of the COVID-19.[10] This encouraged us to accomplish a systematic study on some clinically approved medicines using molecular docking and reinvestigate their biological efficacies and pharmacological properties. However, there is no organized study on the inhibition of the coronavirus by clinically approved drugs to the best of our knowledge. Hence, the present study was aimed to molecular docking studies of clinically approved drugs against the main protease of SARS-CoV-2.

# MATERIALS AND METHODS

### *Molecular Docking Methods*

For molecular docking, Auto-Dock 4.2 software was used.[11] The free energy (DG) binding of SARS-CoV-2 viral protease with the selected FDA approved drugs was created by means of this molecular docking package. Docking is a computational simulation method of a ligand binding to a receptor or enzyme and expects the favored orientation of binding of one molecule to the second to form a steady complex. To predict the attraction and activity of binding of the minor molecule to their enzyme targets by using scoring functions docking is used. Therefore, docking shows significant role in the rational design of medicines. The sensitivity of docking calculations concerning the geometry of the involvement ligand displays that even minor changes in the ligand structure can lead to big changes in the geometries and scores of the subsequent docked poses.

### *Selection of Ligand*

Anti-viral and anti-malarial medications were recognized as potential coronavirus inhibitors from diverse literature evaluations. Total 47 FDA approved drugs were selected for molecular docking with main COVID-19 protease. Among the 47 approved drugs, 2 drugs are anti-H1N1 drugs, 4 are anti-TB drugs, 24 are anti-HIV-1 drugs and 17 are anti-malarial drugs are selected from PubChem database. The three-dimensional structure files of the selected FDA approved drugs were downloaded in SDF format from the PubChem were used for molecular docking. Molecular 2D structures of selected FDA approved drugs are shown in Table 1.

**Table 1**

***Structures of clinically approved drugs***

| | | |
|---|---|---|
| **Oseltamivir (anti-H1N1)** | **Zanamivir (anti-H1N1)** | **Isoniazid (anti-TB)** |
| 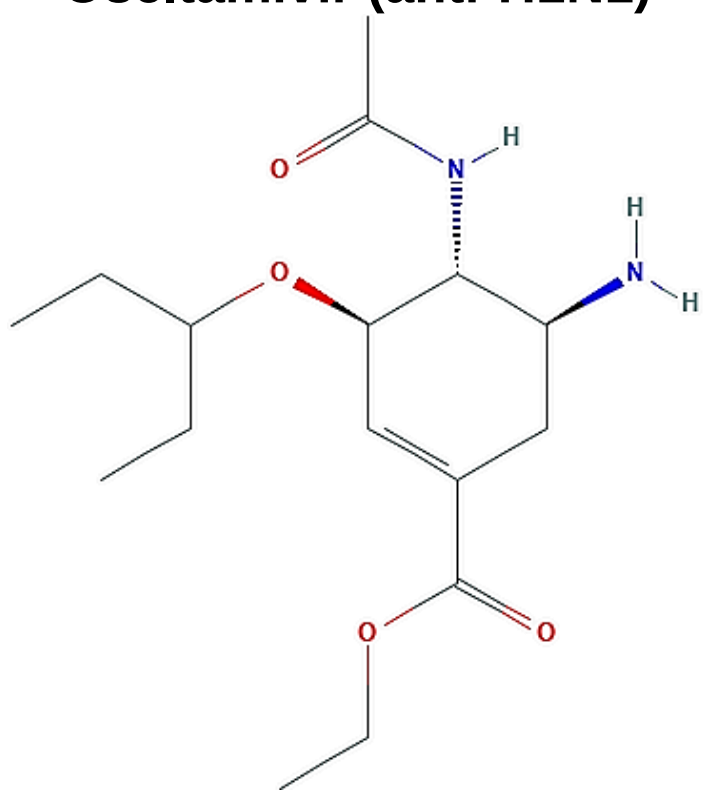 | 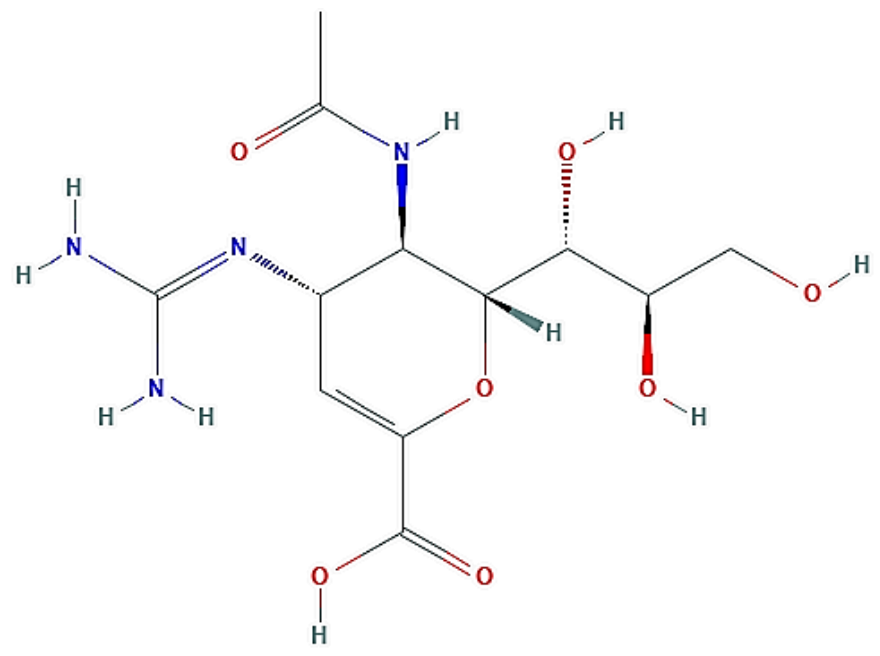 | 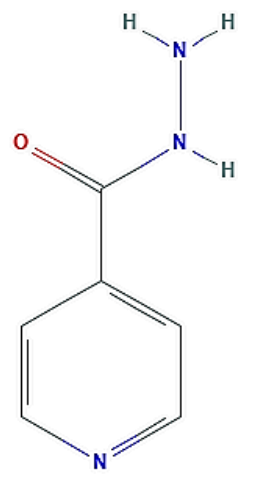 |
| **Rifampin (anti-TB)** | **Ethambutol (anti-TB)** | **Pyrazinamide (anti-TB)** |
| 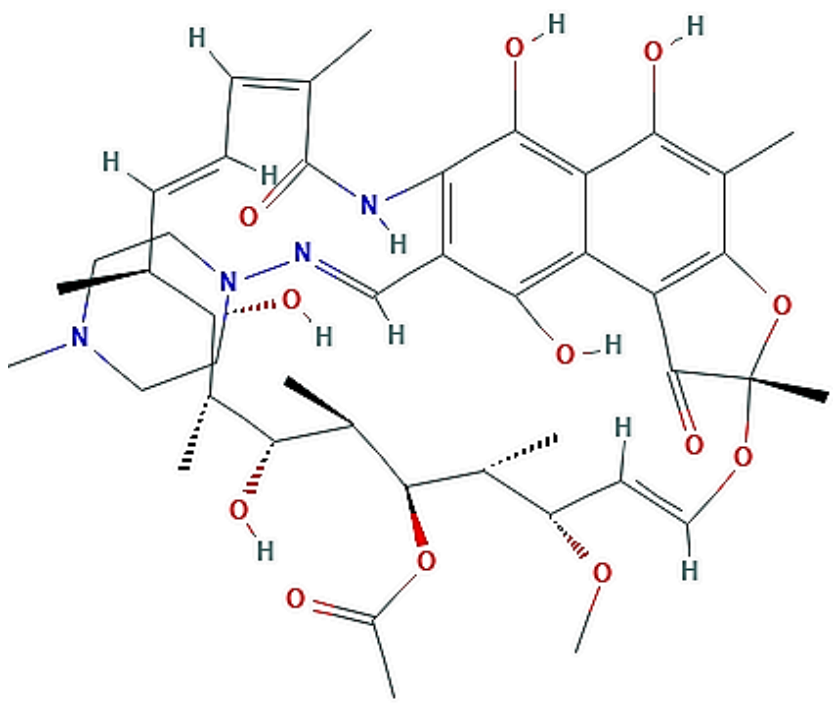 | 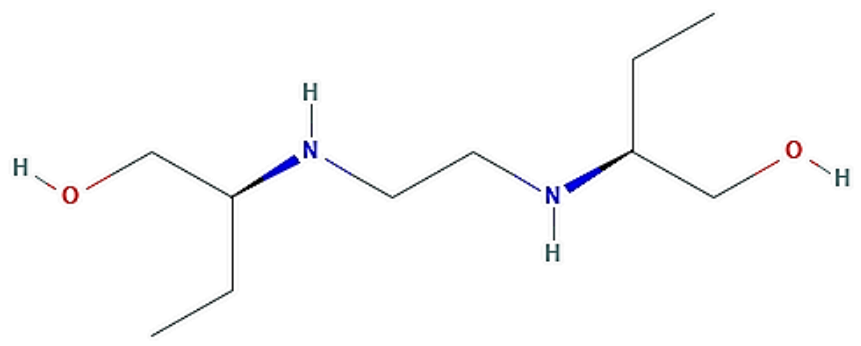 | 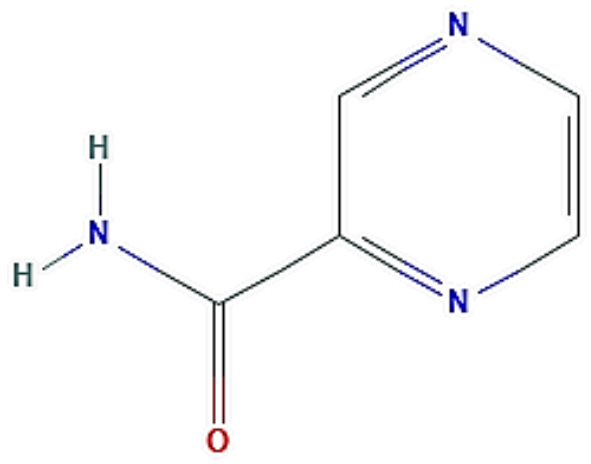 |
| **Abacavir (anti-HIV)** | **Atazanavir (anti-HIV)** | **Darunavir (anti-HIV)** |
| 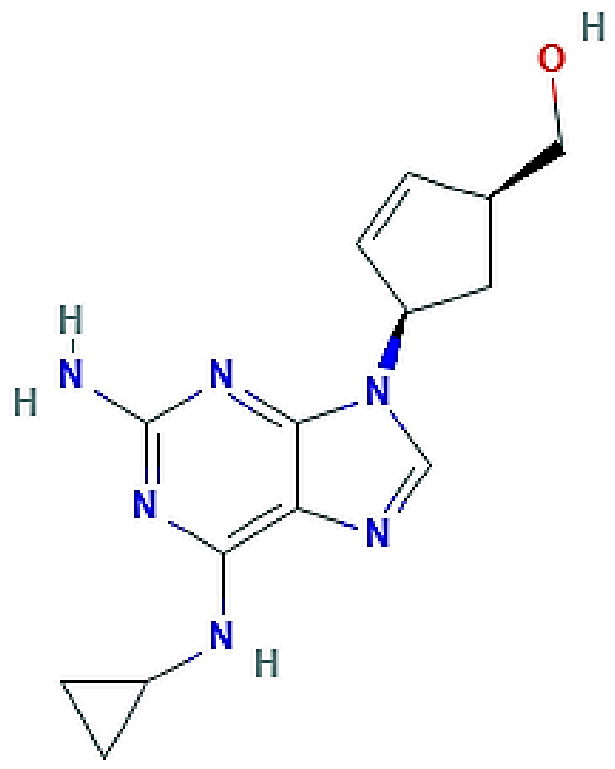 | 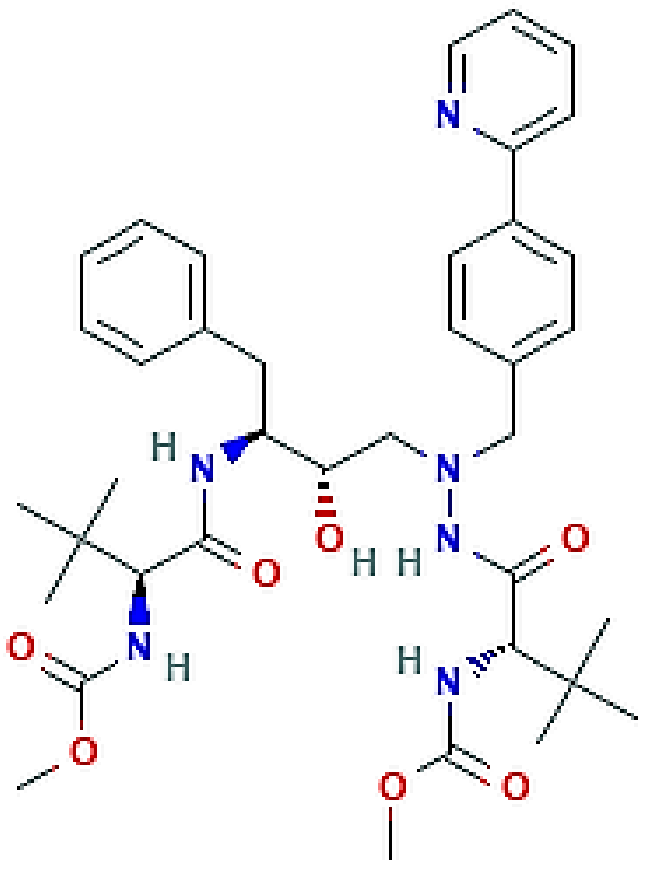 | 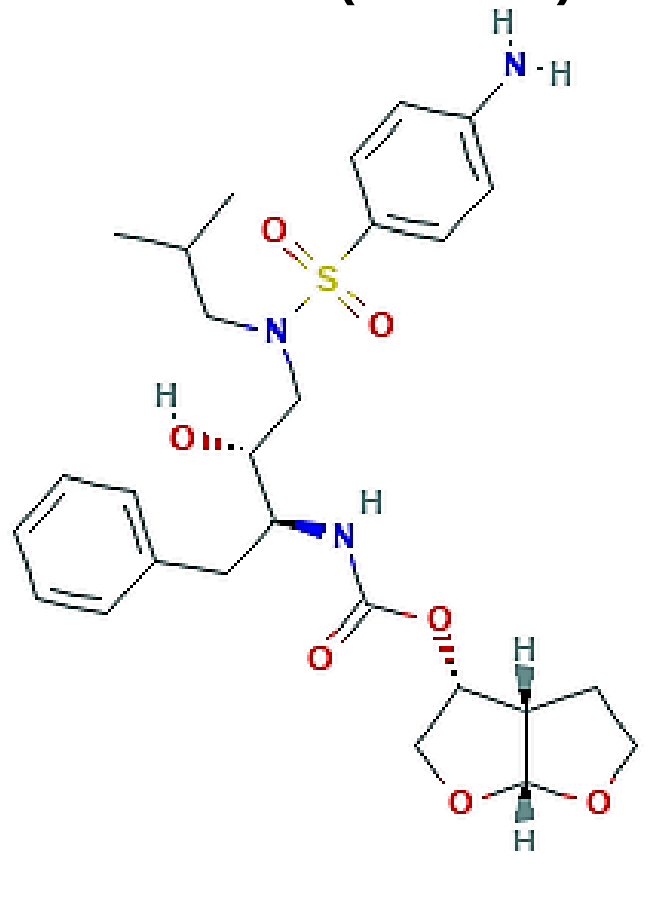 |

**Delavirdine (anti-HIV)**

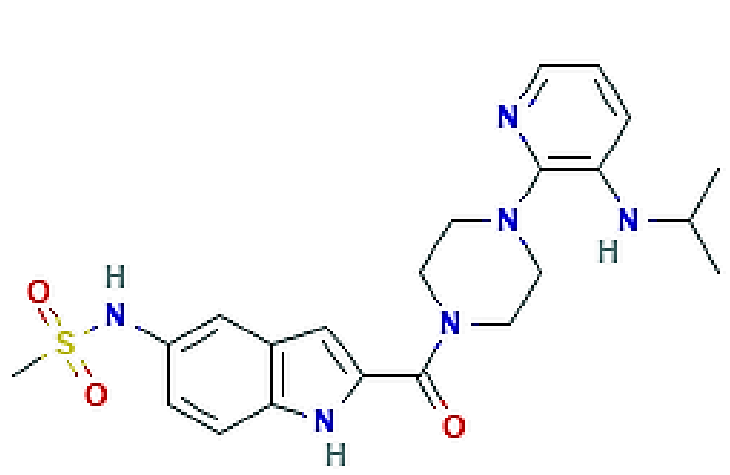

**Dolutegravir (anti-HIV)**

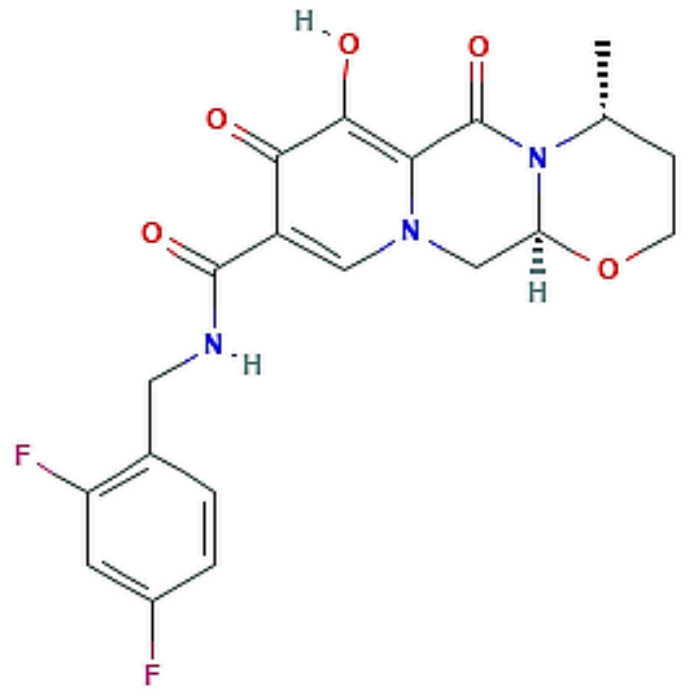

**Doravirine (anti-HIV)**

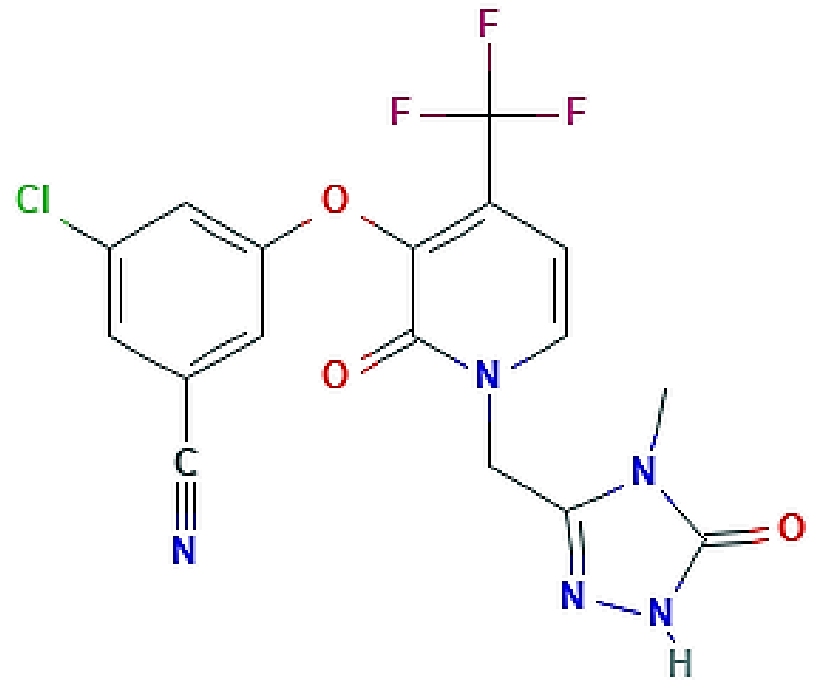

**Efavirenz (anti-HIV)**

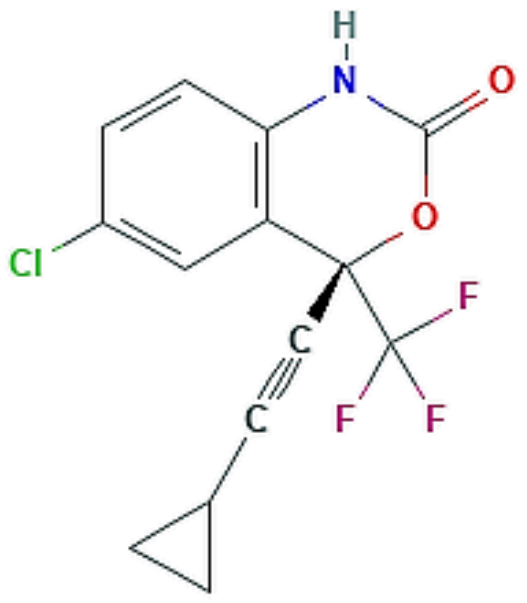

**Elvitegravir (anti-HIV)**

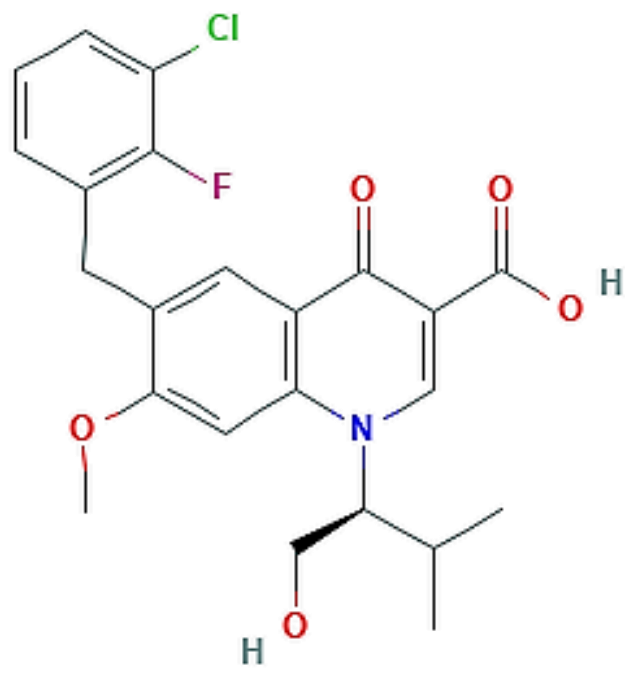

**Emtricitabine (anti-HIV)**

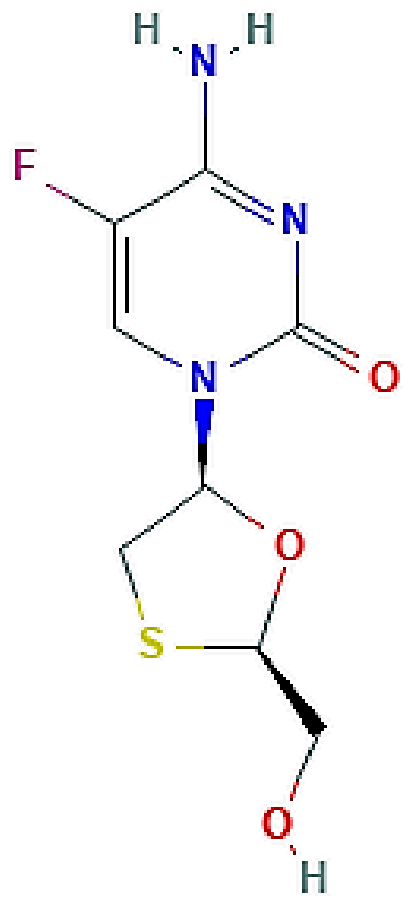

**Etravirine (anti-HIV)**

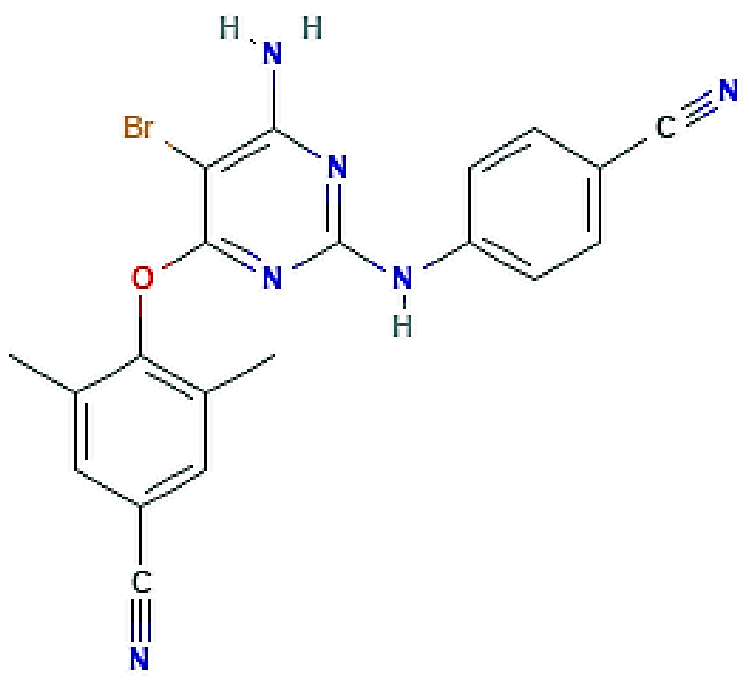

**Fosamprenavir (anti-HIV)**

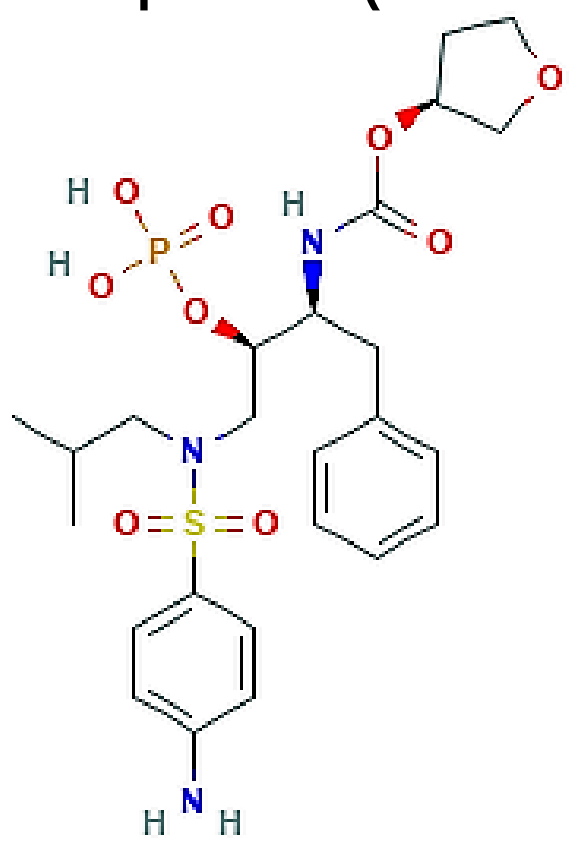

**Indinavir (anti-HIV)**

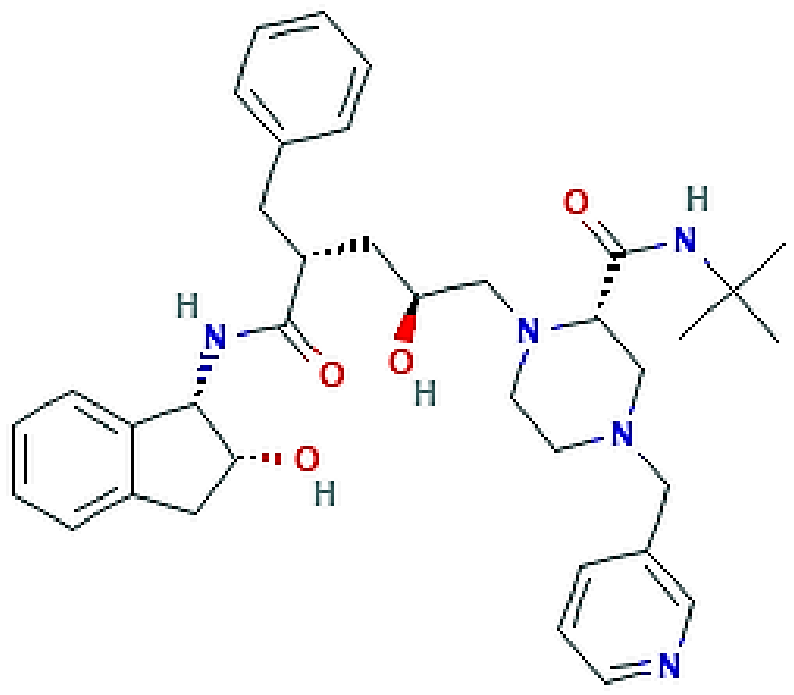

**Lamivudine (anti-HIV)**

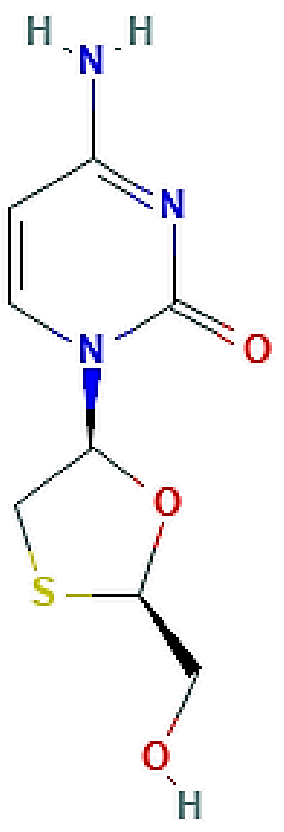

**Lopinavir (anti-HIV)**

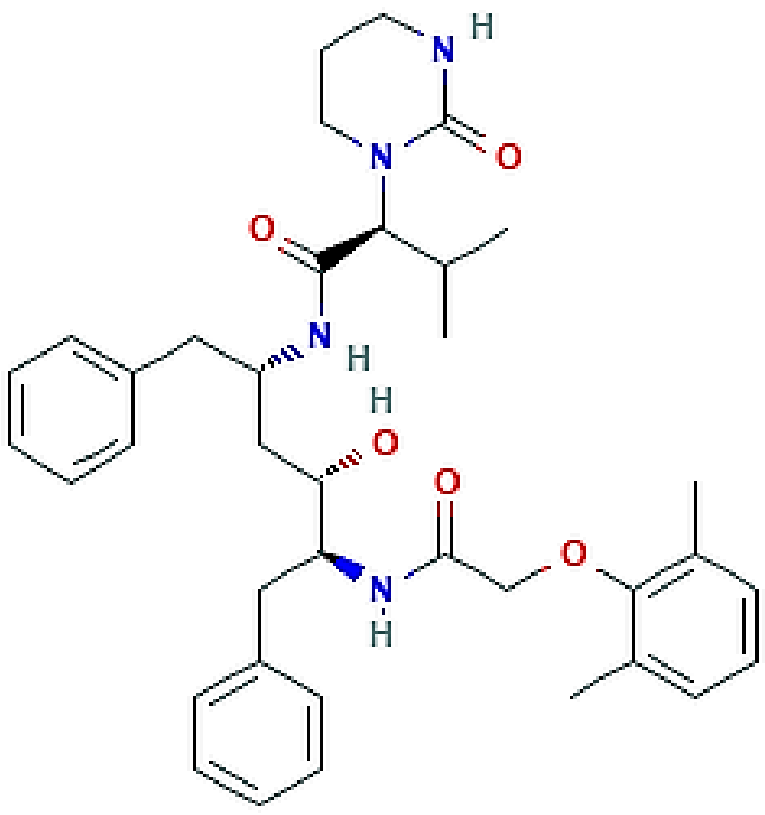

**Maraviroc (anti-HIV)**

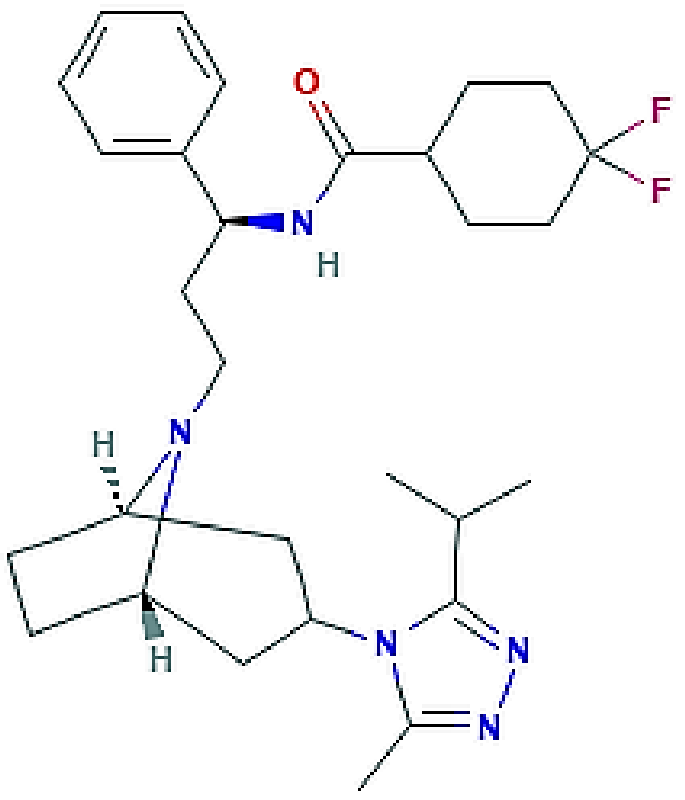

**Nevirapine (anti-HIV)**

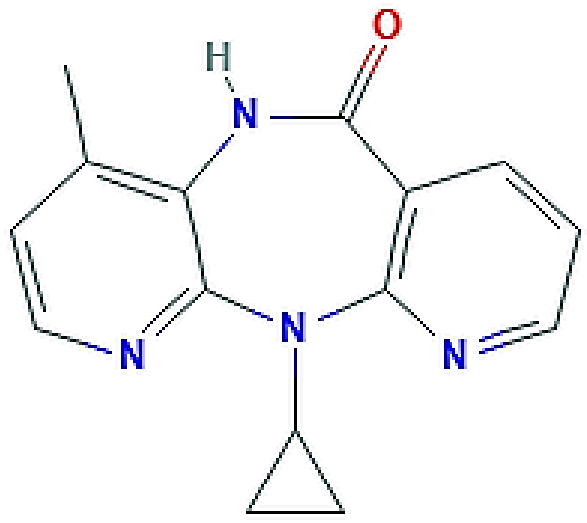

**Raltegravir (anti-HIV)**

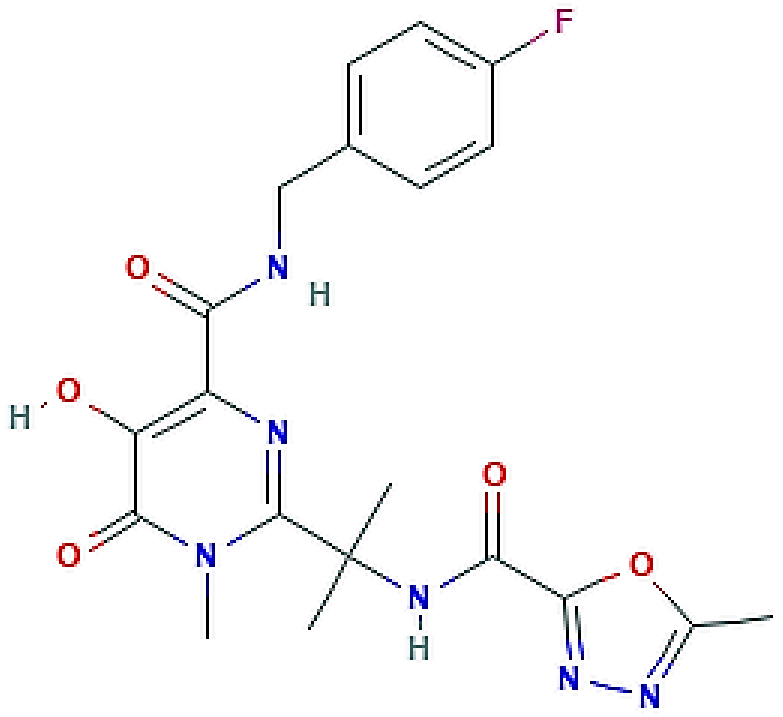

**Rilpivirine (anti-HIV)**

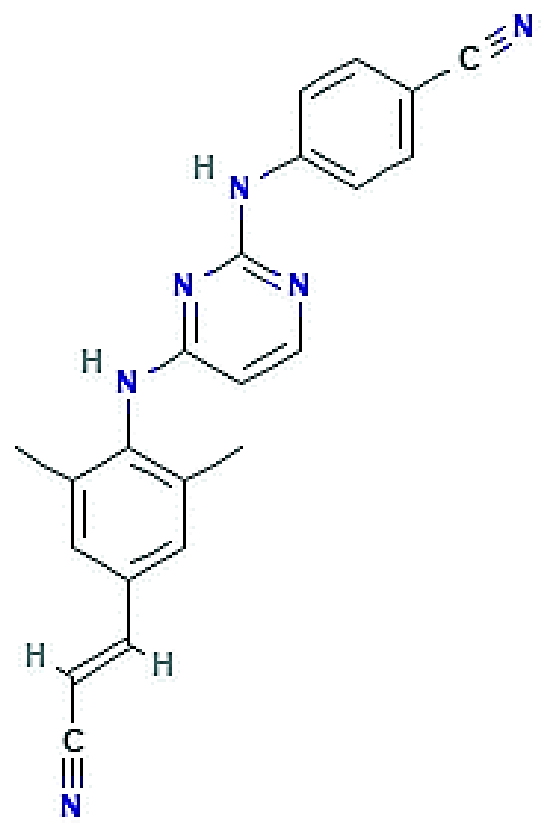

**Ritonavir (anti-HIV)**

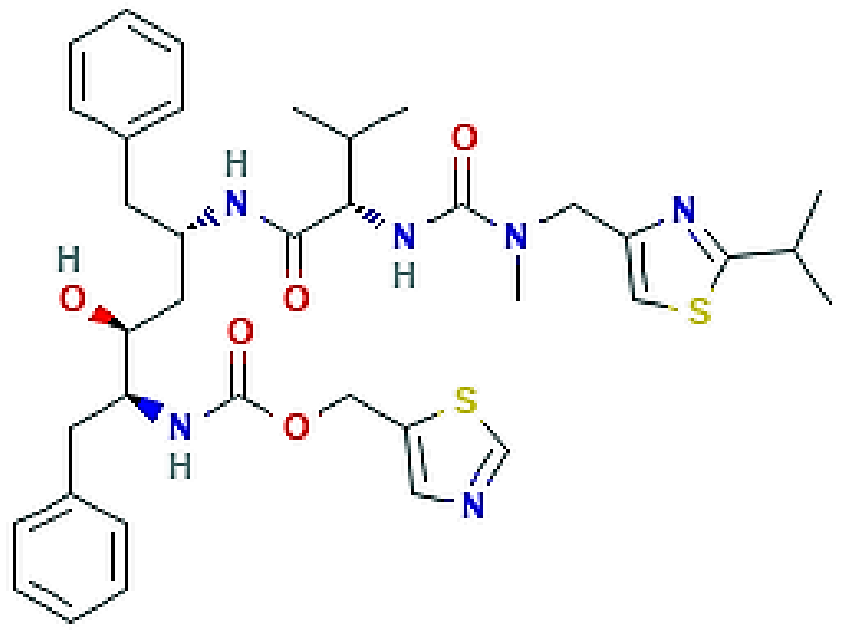

**Saquinavir (anti-HIV)**

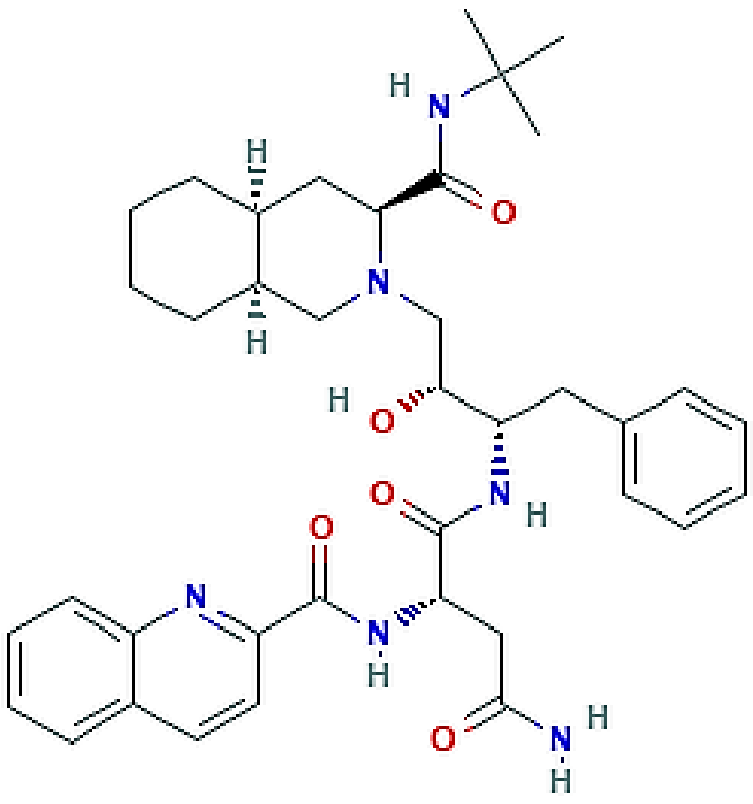

**Stavudine (anti-HIV)**

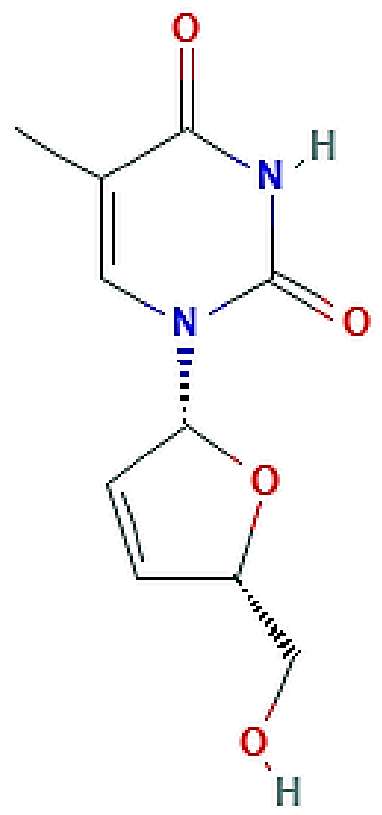

**Tenofovir (anti-HIV)**

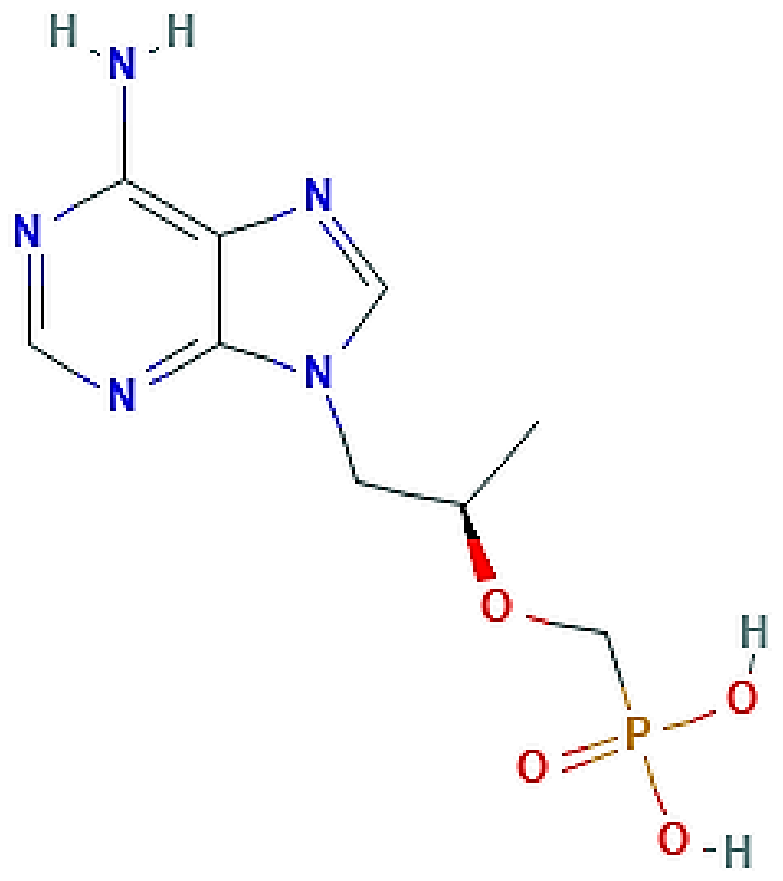

**Tipranavir (anti-HIV)**

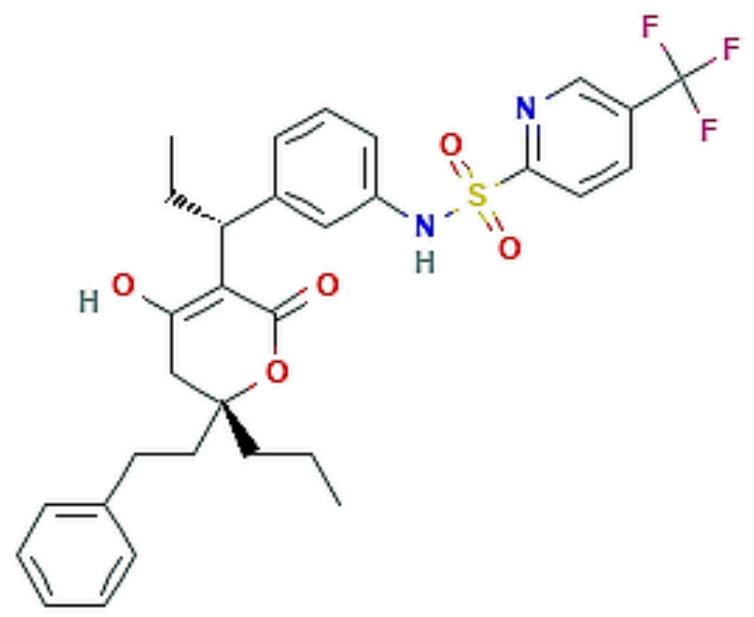

**Zidovudine (anti-HIV)**

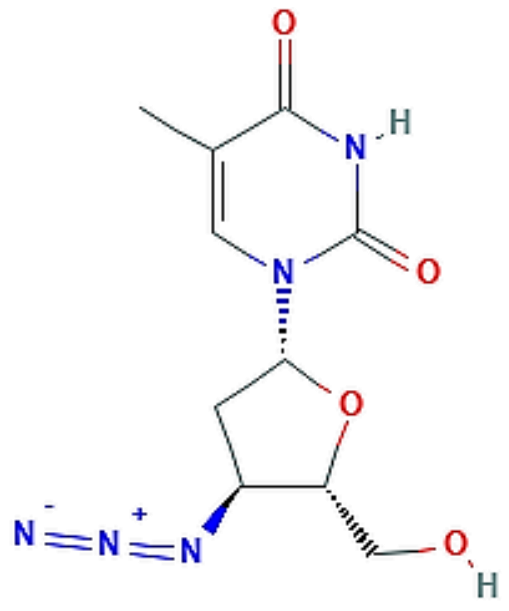

**Quinine (anti-malarial)**

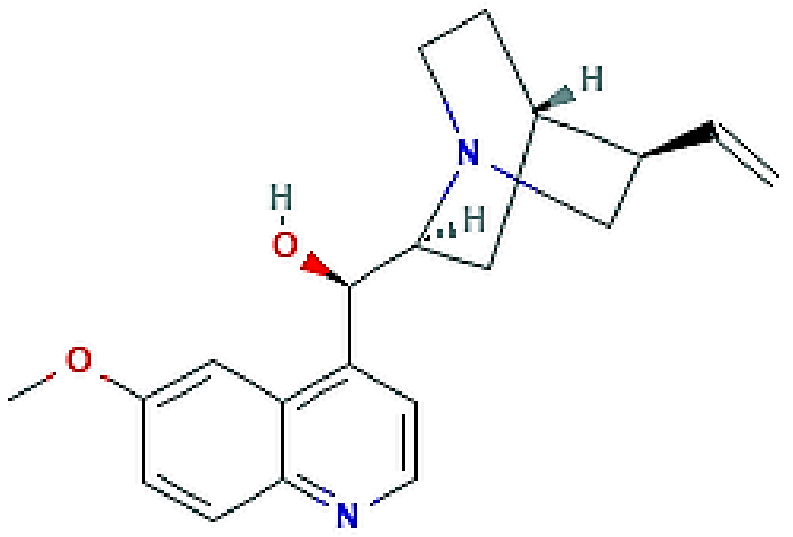

**Quinidine (anti-malarial)**

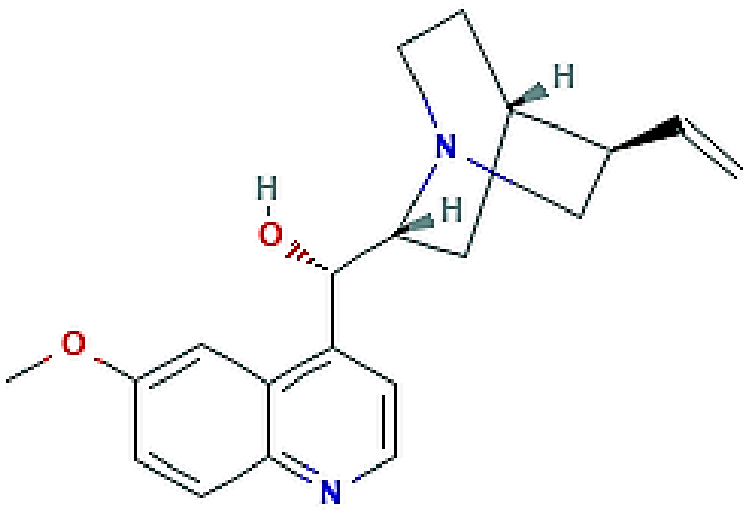

**Mefloquine (anti-malarial)**

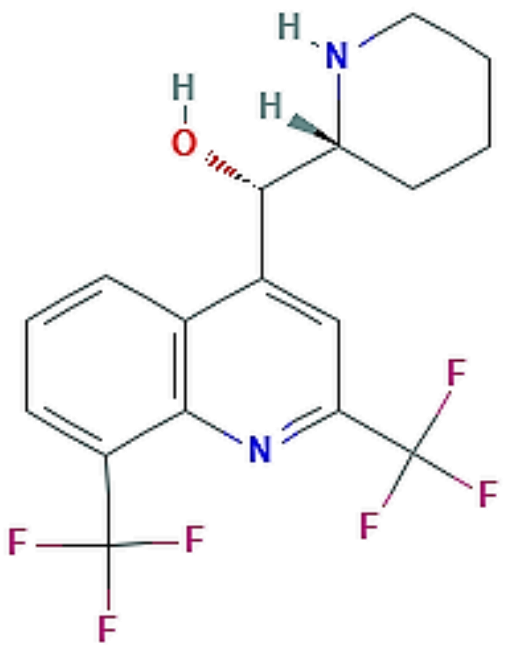

**Chloroquine (anti-malarial)**

**Amodiaquine (anti-malarial)**

**Primaquine (anti-malarial)**

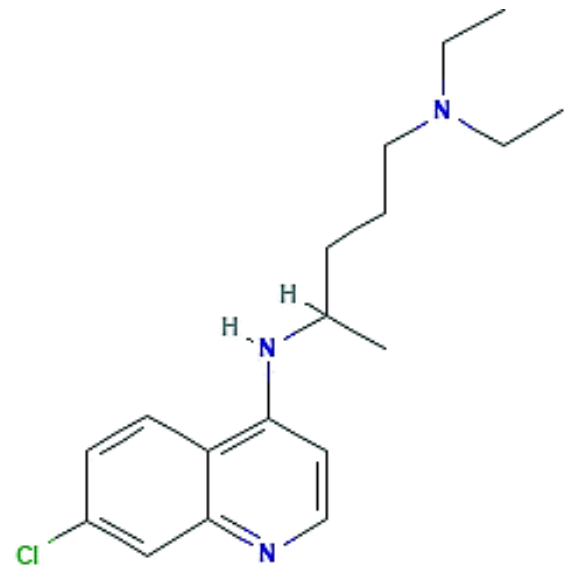

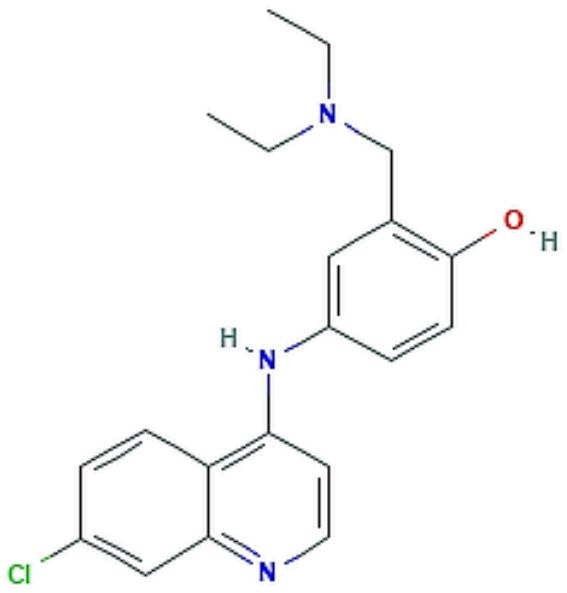

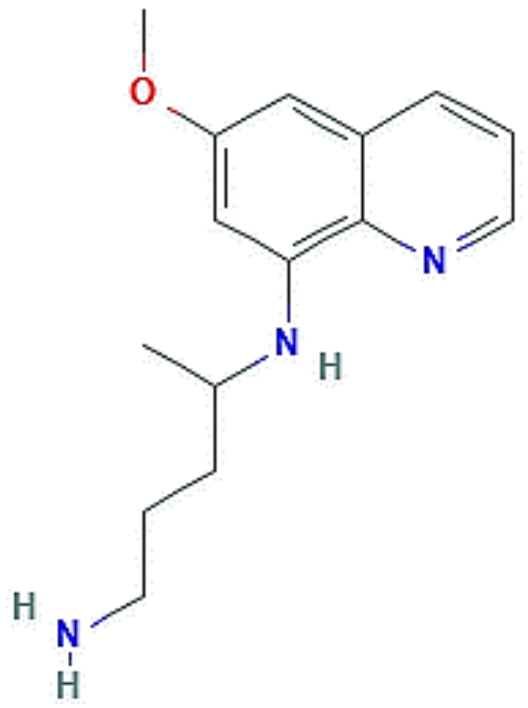

**Halofantrine (anti-malarial)**

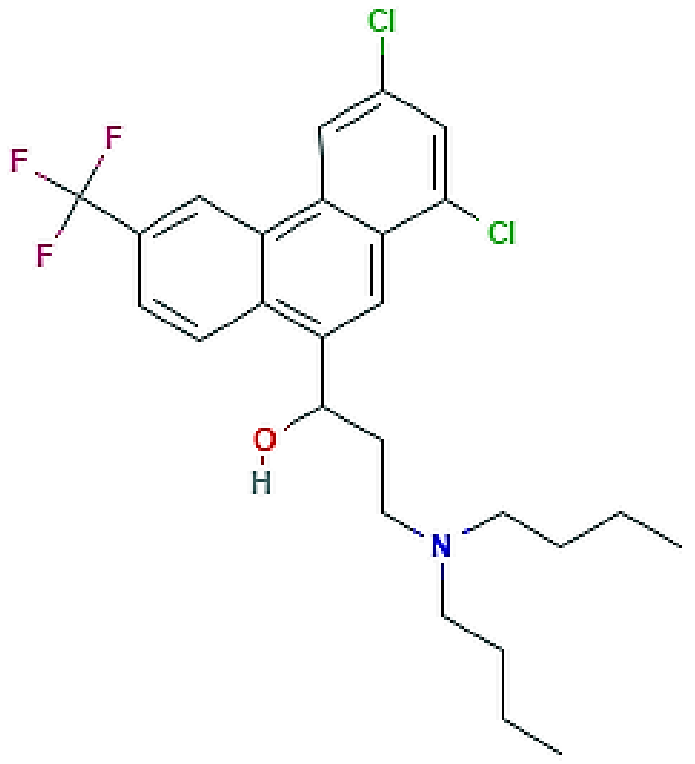

**Sulfadoxine (anti-malarial)**

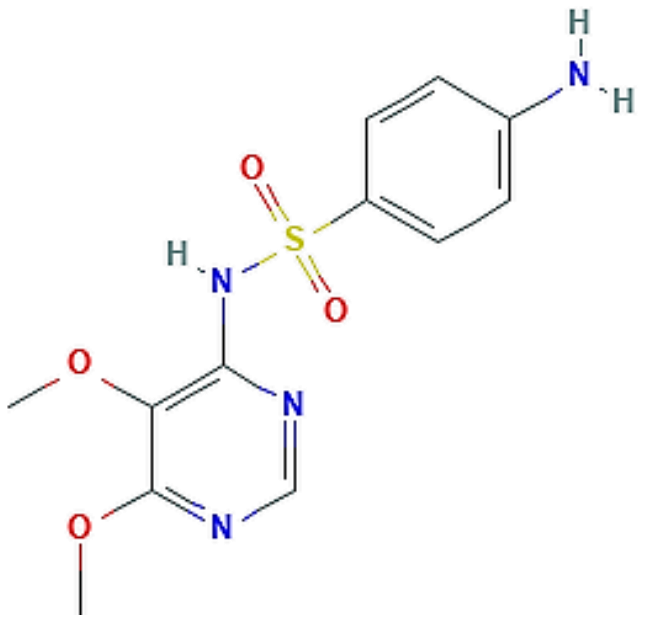

**Sulfamethoxypyridazine (anti-malarial)**

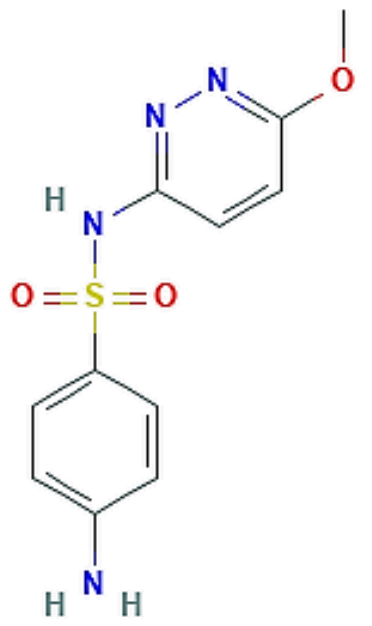

**Proguanil (anti-malarial)**

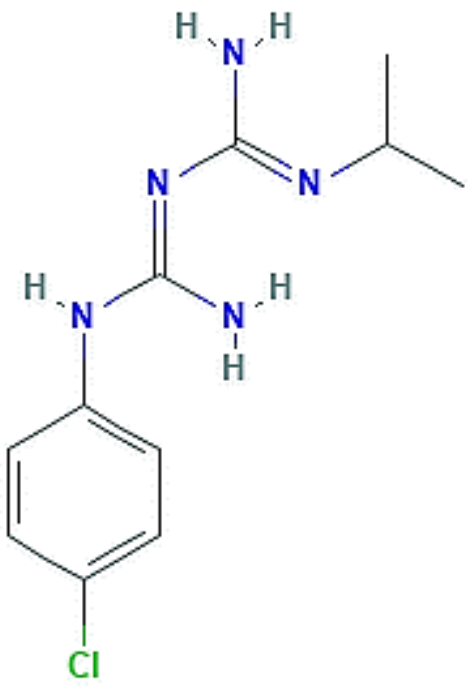

**Pyrimethamine (anti-malarial)**

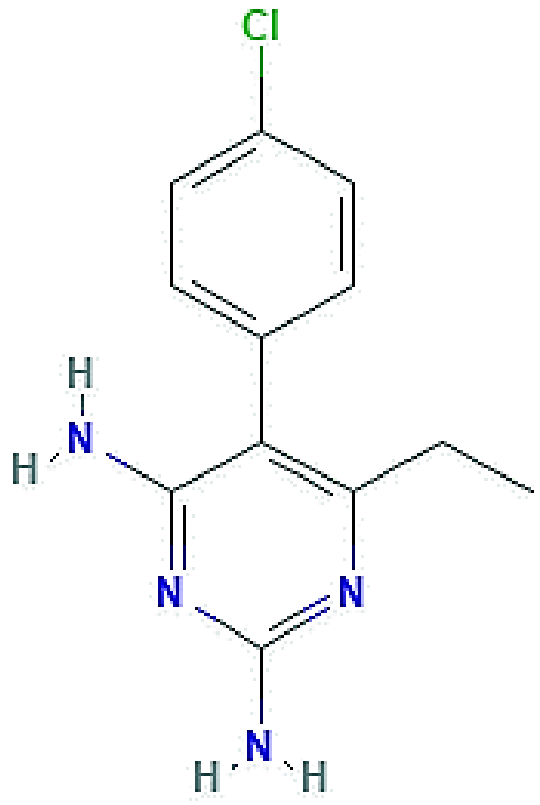

**Tetracycline (anti-malarial)**

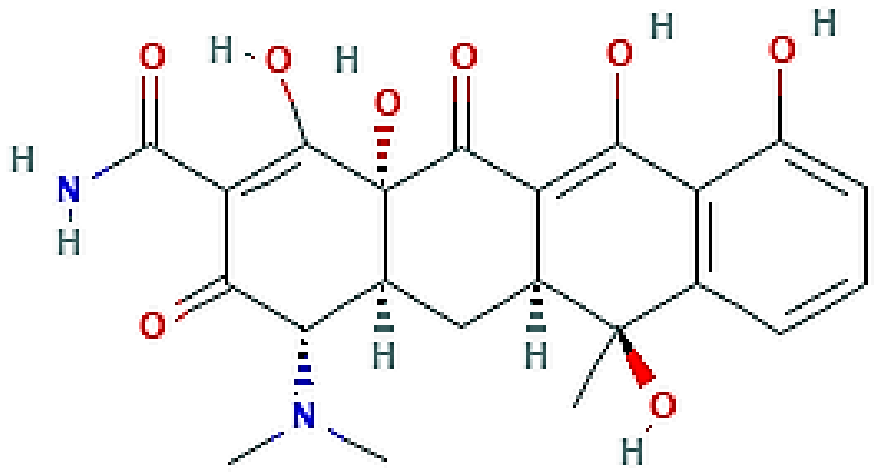

**Doxycycline (anti-malarial)**

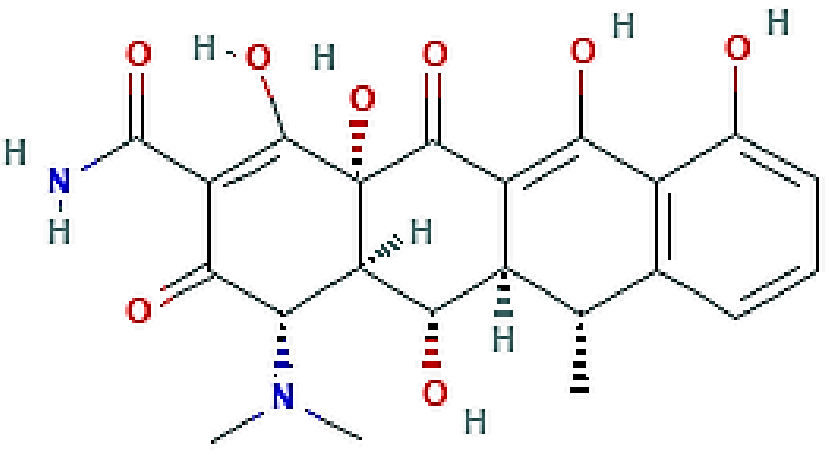

**Clindamycin (anti-malarial)**

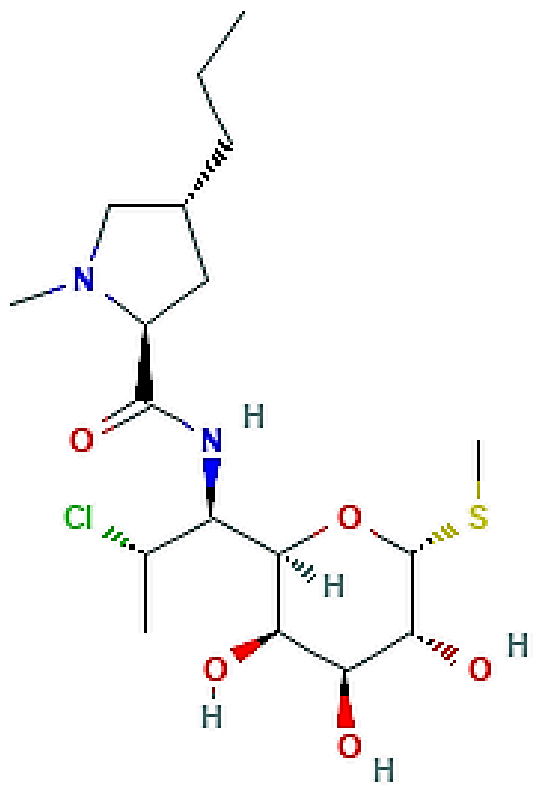

**Azithromycin (anti-malarial)**

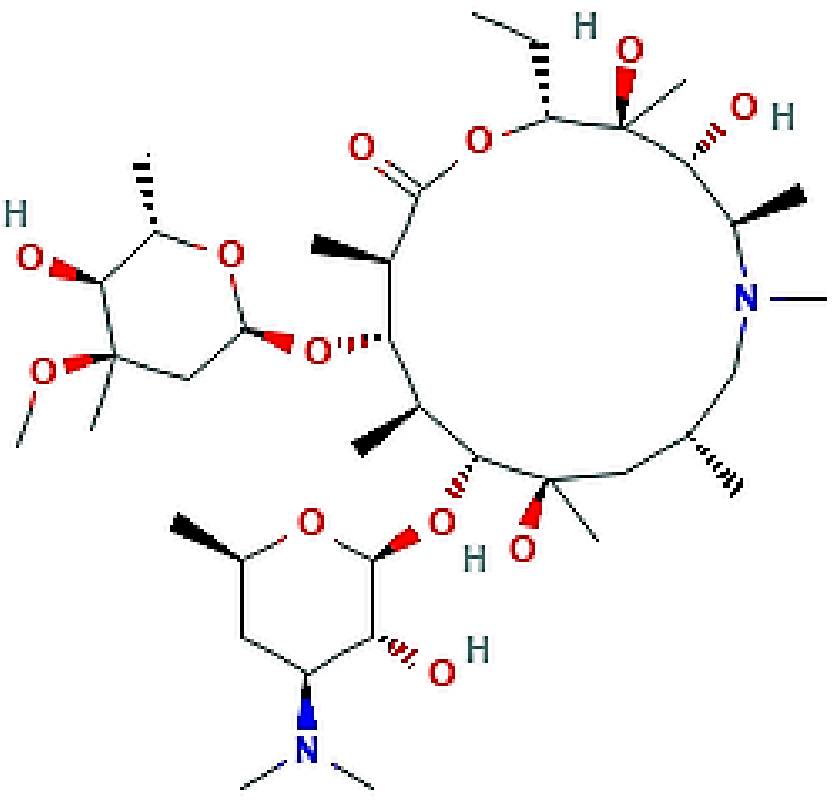

**Artemisinin (anti-malarial)**

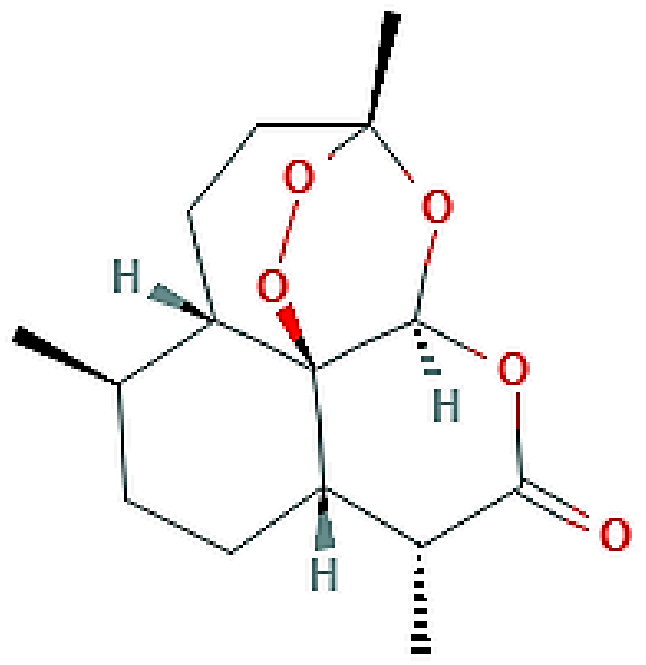

**Atovaquone (anti-malarial)**

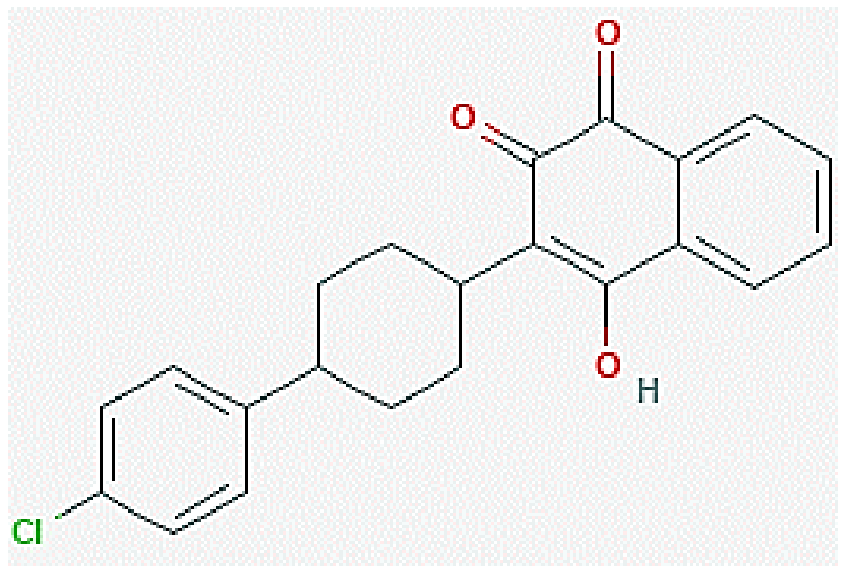

**Hydroxychloroquine (anti-malarial)**

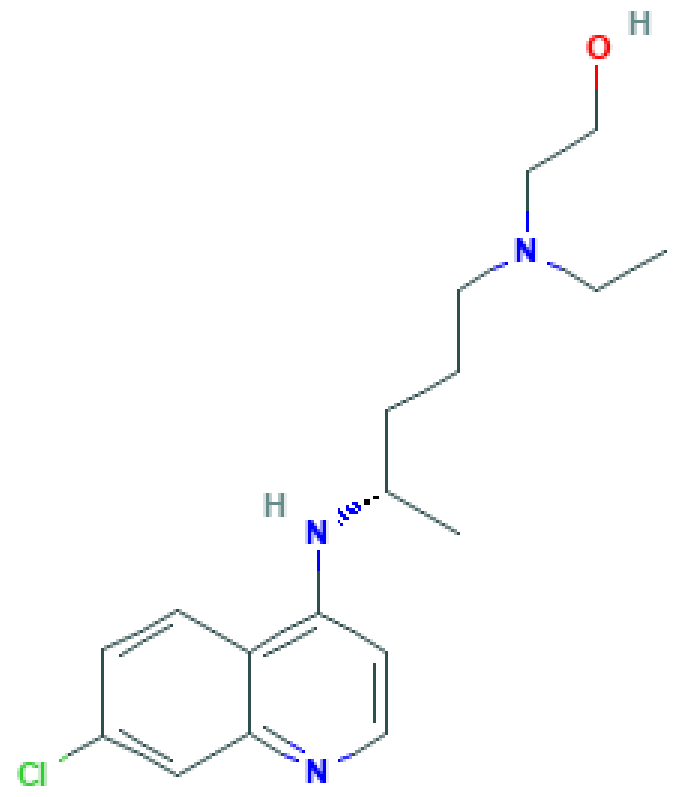

***Selection of Target***

The main COVID-19 protease remained used as a target to novelty repurposing candidates over computational selection amongst clinically accepted drugs. The study identified a list of FDA permitted 47 drugs that may form hydrogen bonds to key residues of amino acids within the binding pocket of viral protease and may too have a higher tolerance to conflict mutations. The crystal 3D structure of SARS-CoV-i2 protease (PDB ID:6LU7) remained obtained from Protein-Data Bank.[12] (**Figure-1**).

**Figure 1**

***The 3D crystal structure of SARS-CoV-2 protease (PDB ID: 6LU7)***

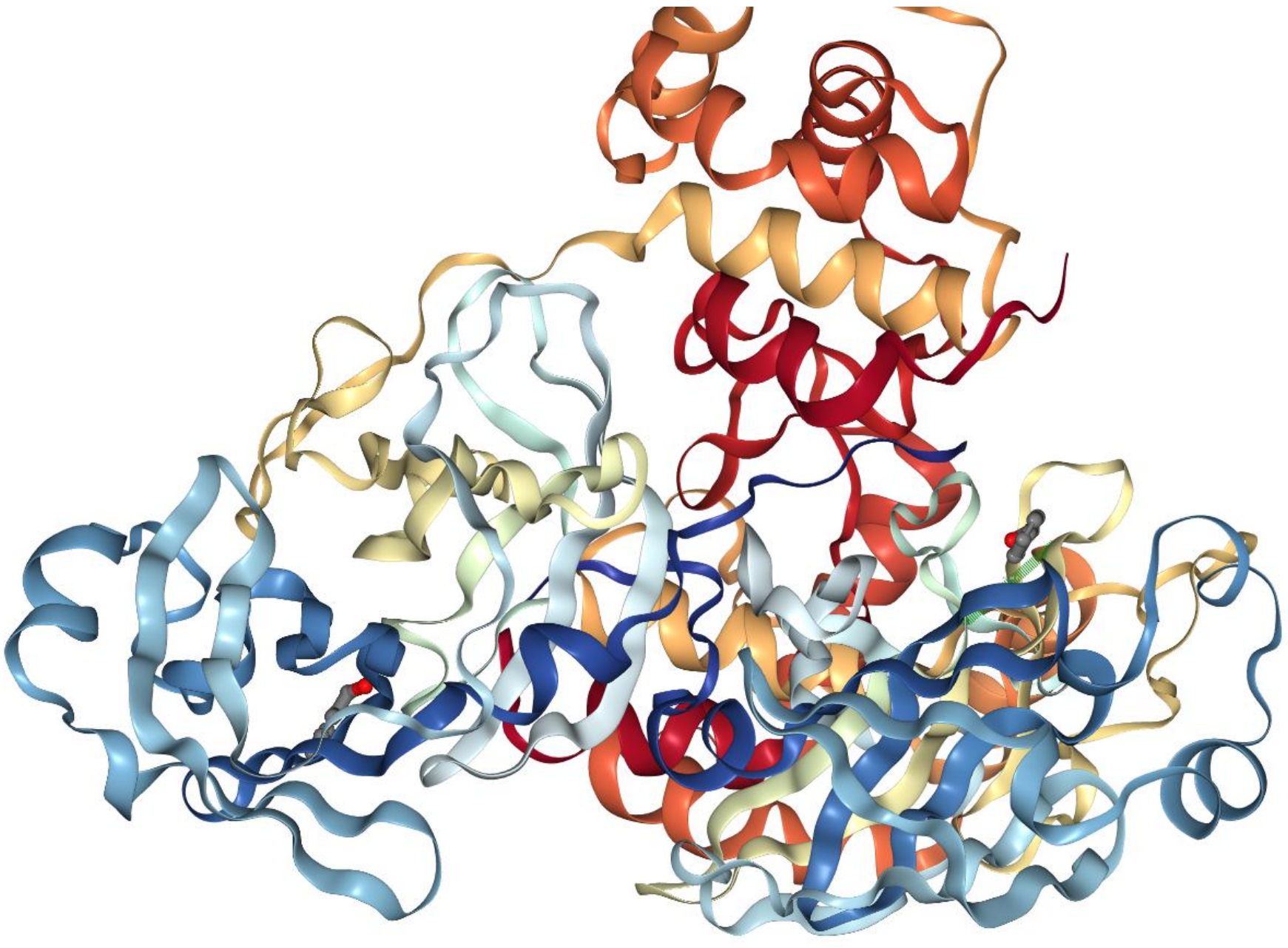

Meanwhile this protease has its crystal structure in a state that signifies the pharmacological target for the progress of new medicines to treatment diverse infectious diseases. The preparation of the target enzyme 6LU7 with the Auto-Dock Tools software intricate addition of all H2 atoms to the enzyme, which is a step essential for accurate calculation of fractional atomic charges. The ligand and all water molecules were detached to make the structure for docking. Gasteiger charges are considered for each atom of the protein in AutoDock 4.2 instead of Kollman charges, which were used in the earlier versions of this package.

***Docking Procedure***

For ligand conformational incisive, we take the 'Lamarckian-genetic algorithm (LGA)', which is a mixture of a genetic algorithm and a native search algorithm. This algorithm initially builds a population of entities, being a diverse casual conformation of the docked enzyme. Each distinct protein is then mutated to attain a slightly diverse translation and alternation and the local search algorithm then achieves energy minimizations on a user-specified amount of the population of individuals. The entities with the low subsequent energy are moved to the succeeding generation and the procedure is then repetitive. This algorithm is called Lamarckian while every novel group of entities is allowable to receive the local search variations of their parents.

To get many docked structures, Auto-Dock was run numerous times, and used to examine the expected docking energy. Rapid energy assessment was attained by pre-calculating nuclear affinity capacities for every atom in the compound molecule. The binding sites of the target enzyme for these molecules in the AutoGrid process were designated on the patterns of founded ligand-binding pockets.[13] Auto-Dock Tools deliver various approaches to examine the outcomes of docking-simulations such as, structural resemblance, and other limitations like inter-molecular energy, visualizing the binding site and its energy and inhibition constant. The energy of interaction of every atom in the ligand was met. For each ligand, 10 best postures were made and scored using Auto-Dock 4.2 scoring purposes.[14]

## RESULTS

Computational approaches for drug discovery and development are proven to be effective and time efficient, as they are not based on difficult laborious works. The protein-ligand docking elucidates the mechanism of inhibition along with the specificity and efficiency of that ligand as an inhibitor. The association of drug candidate (ligand) to its target receptor is a fundamental binding reaction and the aim of the computer-aided drug discovery is to find small molecules having strong inhibitory or activating action against the biological targets. The strength of inhibition or activation is elucidated through binding affinity.[15]

### *Docking Prediction of anti-H1N1 drugs*

Oseltamivir and Zanamivir, two FDA approved drugs docked with SARS-CoV-2 main protease and obtained binding energy is −7.39 kcal/mol and -3.88 kcal/mol respectively (Table-2). Oseltamivir interacted with Glu:166, Pro:52, 168, Met:49,165, Leu:167, His:164, 41, Tyr:54, Gln:189, Arg:188, Asp:187, Thr:190 and Gln:192 at the binding site of this SARS-CoV-2 protease and Zanamivir interacted with Glu:166, Leu:167, Met:165, Gln:189, 192, Thr:190, Ala:191, Pro:168, Gly:170 at the binding site of this protease (Figure 2). The results identified that Oseltamivir is potential inhibitor of the SARS-CoV-2 main protease. Earlier one study has reported that Oseltamivir is a prodrug of oseltamivir carboxylate, a potent and selective inhibitor of the neuraminidase glycoprotein essential for replication of influenza A and B viruses.[16]

**Table 2**
***Molecular docking analysis of anti-H1N1 drugs against COVID-19 Protease (6LU7)***

| Sl. N | Compound Name | Binding energy (kcal/mol) | Residue involving interaction | No. of H bonds | Interaction of residues forming $H_2$ bonds |
|---|---|---|---|---|---|
| 1 | Oseltamivir | -7.39 | GLU:166, PRO:52, 168, MET:49,165, LEU:167, HIS:164, 41, TYR:54, GLN:189, ARG:188, ASP:187, THR:190, GLN:192 | 1 | GLU:166 |
| 2 | Zanamivir | -3.88 | GLU:166, LEU:167, MET:165, GLN:189, 192, THR:190, ALA:191, PRO:168, GLY:170 | 4 | GLU:166, LEU:167 |

**Figure 2**
***Docking visualisation of COVID-19 protease (6LU7) with Oseltamivir (anti-H1N1 drug)***

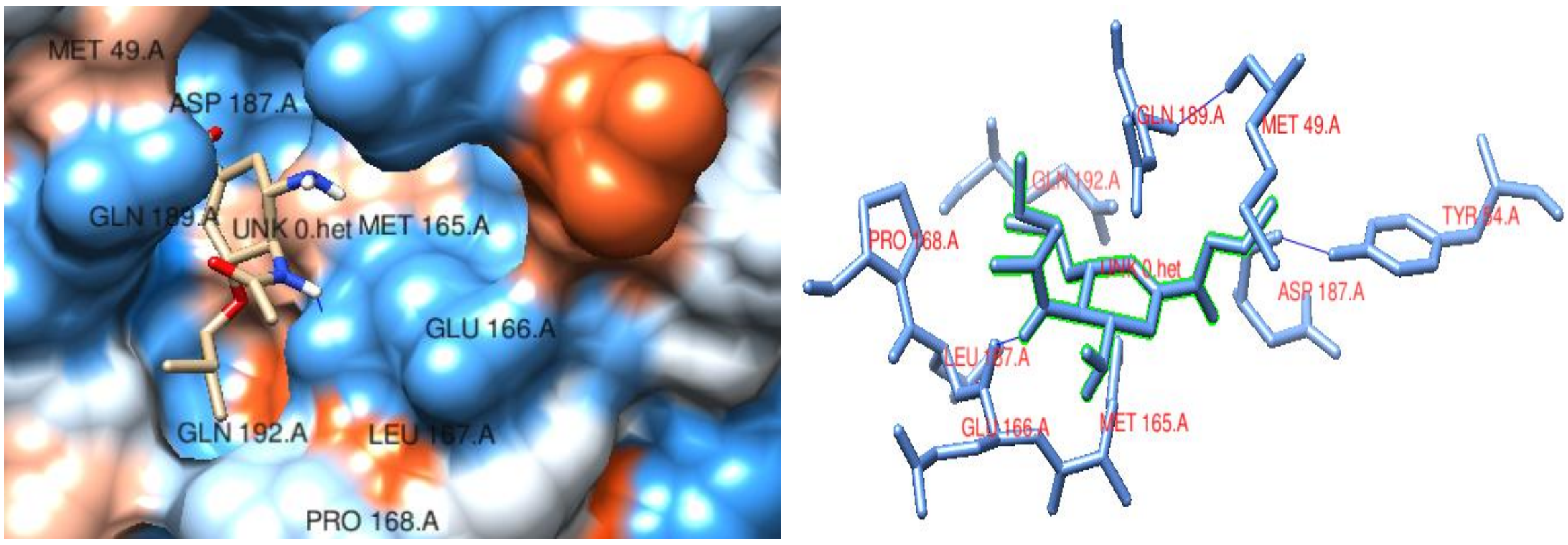

### *Docking Prediction of anti-TB drugs*

Isoniazid, Rifampin, Ethambutol and Pyrazinamide are clinically approved drugs were docked with binding energy -4.83, -9.41, -5.02 and -4.05 kcal/mol respectively against SARS-CoV-2 protease (Table 3). Rifampin showing highest binding affinity -9.41 kcal/mol among all the four drugs. The residues involved in the interaction with the Rifampin were Glu:166, Met:165, His:163, 172, 164, 163, Phe:140, Leu:141, 167, Ser:144, Gly:143, Asn;142, Pro:163, Gln:192, Cys:145, Ala:191, Thr:190 and Gln:189 (Figure 3).

**Table 3**
***Molecular docking analysis of anti-TB drugs against COVID-19 Protease (6LU7)***

| Sl. N | Compound Name | Binding energy (kcal/mol) | Residue involving interaction | No. of H bonds | Interaction of residues forming $H_2$ bonds |
|---|---|---|---|---|---|
| 1 | Isoniazid | -4.83 | MET:49, 165, ASP:187, TYR:54, GLN:189, HIS:41, PRO:52, ARG:188 | 4 | MET:49, TYR:54, ASP:187 |
| 2 | Rifampin | -9.41 | GLU:166, MET:165, HIS:163, 172, 164, 163, PHE:140, LEU:141, 167, SER:144, GLY:143, ASN;142, PRO:163, GLN:192, CYS:145, ALA:191, THR:190, GLN:189 | 1 | GLU:166 |
| 3 | Ethambutol | -5.02 | LEU:141, ASN;142, GLU:166, HIS:163,172, MET:165, SER:144, GLY:143, CYS:145, PHE:140, GLY:170 | 4 | LEU:141, ASN:142, GLU:166 |
| 4 | Pyrazinamide | -4.05 | VAL:303, 212, THR:304, 257, GLN:256, GLN:306, ARG:217, ILE:213 | 3 | VAL:303, THR:304, GLN:256 |

**Figure 3**
***Docking visualisation of COVID-19 protease (6LU7) with anti-TB drug Rifampin***

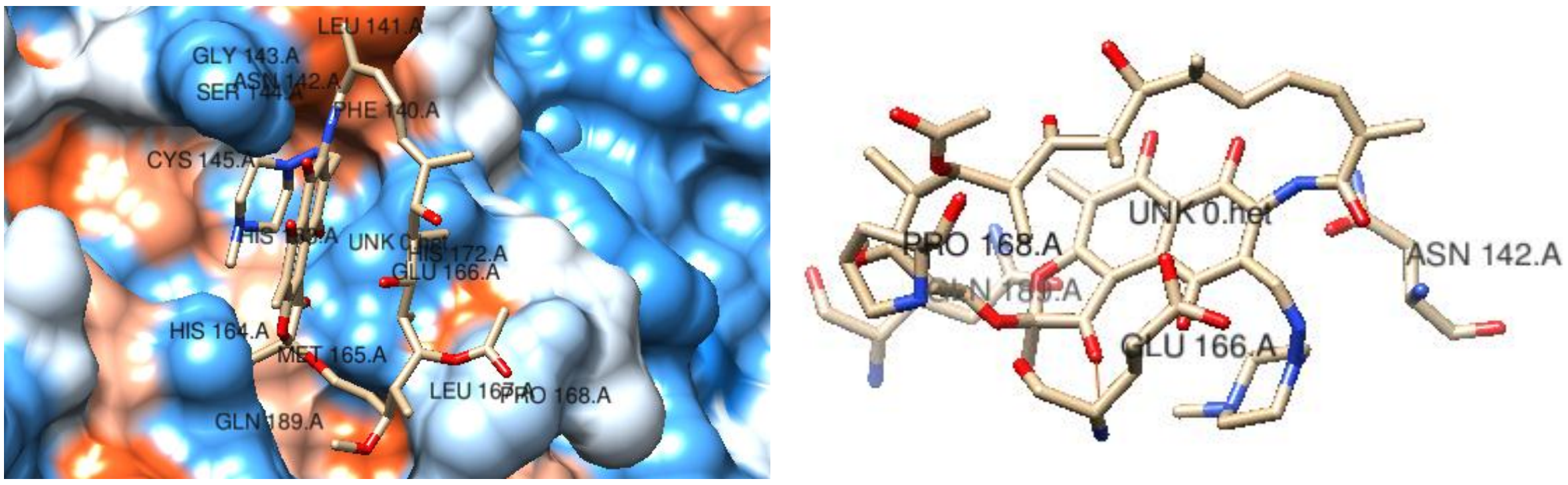

***Docking Prediction of anti-HIV drugs***

Twenty-four FDA approved anti-HIV drugs were docked with SARS-CoV-2 protease. Among the twenty four drugs, four drugs Maraviroc, Etravirine, Indinavir and Rilpivirine were showed more potential inhibitors of SARS-CoV-2 main protease with binding affinity -10.67, -10.33, -10.00 and -9.66 kcal/mol respectively (Table 4). Maraviroc interacted at His:164, 41, Tyr:54, Asp:187, Met:165, 49, Leu:167,141, Pro:168, Cys:44, 145, Arg:188, Gly:143, Asn:142, Ala:191, Gln:192, 189, Thr:190, Etravirine interacted at Gln:83, Lys:88, Thr:175, Met:82, 162, His:164, Cys:85,38, Gly:179, Pro:39, Leu:177,87, Cys:38, Tyr:37, Glu:178, Asn:84, Arg:40, Indinavir interacted at Glu:166, Gln:189, 192, Cys:145, Met:49, 165, Asp:187, His:41, 164, 163, Asn:142, Leu:167, 141, Pro:168, Thr:190, Phe:140, Ser:144, 46, Arg:188 and Rilpivirine interacted at Glu:166, Cys:44, 145, His:164, 41, Met:165, 49, Ala:191, Pro:168,52, Gln:189, Tyr:54, Asp:187, Arg:188, Leu:167, Thr:190 of the SARS-CoV-2 main protease (Figure 4). A joint research team of the Shanghai Institute of Materia Medica and Shanghai Tech University performed drug screening in silicon and an enzyme activity test, and they reported 30 anti-HIV agents with potential antiviral activity against SARS-CoV-2 on January 25, 2020.[17]

**Table 4**
***Molecular docking analysis of anti-HIV drugs against COVID-19 Protease (6LU7)***

| Sl. N | Compound Name | Binding energy (kcal/mol) | Residue involving interaction | No. of H bonds | Interaction of residues forming $H_2$ bonds |
|---|---|---|---|---|---|
| 1 | Abacavir | -7.77 | MET:49, 165, GLN:189,192, HIS:41, THR:190, ASP:189, TYR:54, PRO:52,168 LEU:50,167, ARG:188, GLU:166, | 1 | MET:49 |
| 2 | Atazanavir | -5.08 | LYS:137,5, GLU:288,290, TYR:126, LEU:287,286, ASP:289, VAL:125, PHE:3, 291, ARG:4, CYS:128, GLY:138, GLN:127, ALA:7, | 1 | LYS:137 |
| 3 | Darunavir | -6.08 | GLN:189,192, GLU:166, PHE:140, PRO:168, MET:165, LEU:167, THR:190, GLN:192, GLY:143, CYS:145, SER:144, HIS:163,164, 172, LEU:141, ASN:142, | 3 | GLN:189, GLU:166, PHE:140 |
| 4 | Delavirdine | -7.89 | TYR:54, GLU:166, ALA:191, MET:165,49, LEU:167, 50, GLN:192, 189, PRO:168, HIS:41,164, ARG:188, ASP:187, THR:190, | 2 | TYR:54, GLU:166 |
| 5 | Dolutegravir | -7.75 | GLU:166, MET:165, 49, GLN:192, THR:190, PRO:168, LEU:167, 27 ALA:191, ARG:188, GLN:189, HIS:164, 41, CYS:145, GLY:143, ASN:142, SER:144 | 1 | GLU:166 |
| 6 | Doravirine | -8.15 | THR:190, 25,ARG:188, TYR:54, MET:165, 49, ASP:187, HIS:41, GLN:189, 192, GLU:166, HIS:164, GLY:143, LEU:27, THR:25, | 3 | THR:190, TYR:54, ARG:188 |
| 7 | Efavirenz | -6.61 | GLY:143, CYS:145, ASN:142, MET:49, HIS:41,163, 164, THR:26, LEU:27, 141, SER:144, PHE:140, GLU:166, MET:165, GLN:189 | 1 | GLY:143 |
| 8 | Elvitegravir | -7.98 | THR:190,ASN:142, CYS:145, MET:165, 49, GLN:192, 189, PRO:168, ARG:188, GLU:166, LEU:167, 141, GLY:143, SER:144, HIS:163, 164, PHE:140, ASP:187, | 1 | THR:190 |
| 9 | Emtricitabine | -4.79 | HIS:41, 164, GLU:166, MET:165,49, THR:190, ARG:188, GLN:192, 189, TYR:54, ASP:187, | 2 | HIS:41, GLU:166 |
| 10 | Etravirine | -10.33 | GLN:83, LYS:88, THR:175, MET:82, 162, HIS:164, CYS:85,38, GLY:179, PRO:39, LEU:177,87, CYS:38, TYR:37, GLU:178, ASN:84, ARG:40 | 2 | GLN:83 |
| 11 | Fosamprenavir | -4.08 | ASN:203, 151, PRO:293, PHE:8,294, ILE:249, 200, 106, VAL:202, 104, GLN:110, THR:292, GLY:109, ASP:153, VAL:104, SER:158, | 1 | ASN:203 |
| 12 | Indinavir | -10.00 | GLU:166, GLN:189, 192, CYS:145, MET:49, 165, ASP:187, HIS:41, 164, 163, ASN:142, LEU:167, 141, PRO:168, THR:190, PHE:140, SER:144, 46, ARG:188 | 3 | GLU:166, GLN:189 |
| 13 | lamivudine | -4.75 | PHE:140, GLU:166, SER:144, CYS:145, HIS:164, 41, 163, 172, MET:49, 165, ASN:142,LEU:141, GLY:143 | 5 | PHE:140, GLU:166, SER:144, CYS:145, HIS:164 |
| 14 | Lopinavir | -6.11 | GLU:166, ASN:142, SER:144, HIS:163,164 PHE:140, CYS:145, LEU:141,167 ARG:188, MET:165, GLN:189,192, ALA:191, THR:190 | 1 | GLU:166 |
| 15 | Maraviroc | -10.67 | HIS:164, 41, TYR:54, ASP:187, MET:165, 49, LEU:167,141, PRO:168, CYS:44, 145, ARG:188, GLY:143, ASN:142, ALA:191, GLN:192, 189, THR:190, | 2 | TYR:54, HIS:164, |
| 16 | Nevirapine | -6.44 | MET:49, 165, HIS:41, 164, SER:144, CYS:145, GLU:166, GLN:189, ASP:187, ARG:188, TYR:54 | 0 | 0 |
| 17 | Raltegravir | -7.81 | SER:144, CYS:145, GLU:166, PRO:168,MET:49, 165, HIS:41,163,172, 164, LEU:141, 167, PHE:140, GLY:143, ASN:142, GLN:189, THR:190, ARG:188, | 3 | SER:144, CYS:145, GLU:166 |
| 18 | Rilpivirine | -9.66 | GLU:166, CYS:44, 145, HIS:164, 41, MET:165, 49, ALA:191, PRO:168,52, GLN:189, TYR:54, ASP:187, ARG:188, LEU:167, THR:190, | 3 | HIS:164, CYS:44, GLU:166 |
| 19 | Ritonavir | -8.25 | GLY:143,170, GLU:166, LEU:141,27,167, PHE:140, CYS:145, ASN:142, HIS:41, ASP:187, MET:49,163 ARG: 188, GLN:189,192, THR:190, ALA:191, GLY:170, PRO:168, | 1 | GLY:143 |
| 20 | Saquinavir | -7.55 | GLY:302, 2, ARG:298, SER:301,1, PRO:9, VAL:297,303, MET:6, CYS:300, ALA:7, PHE:8, GLY:2 | 5 | ARG:298, GLY:302, SER:301 |
| 21 | Stavudine | -5.38 | ARG:188, GLN:189, 192, HIS:41, 164, MET:165, THR:190, GLN:192, VAL:186, TYR:54, ASP:187, GLU:166 | 2 | ARG:188 |
| 22 | Tenofovir | -4.40 | THR:190, GLU:166, MET:165, 49, ALA:191, GLN:189, 192, ARG:188, ASP:187, PRO:168, LEU:167, HIS:41 | 2 | THR:190, GLU:166 |
| 23 | Tipranavir | -8.30 | GLN:189, GLU:166, ASN:142, MET:49, CYS:145, MET:165, 49, HIS:41, 164, 163, SER:144, 46, PHE:140, LEU:141, ARG:188, ASP:187, TYR:54, GLY:143, | 3 | GLN:189, ASNI142, GLU:166 |
| 24 | Zidovudine | -6.31 | ASP:187, GLU:166, ARG:188, TYR:54, MET:165,49, GLN:189, HIS:164,41, PHE:181, PRO:52CYS:44 | 2 | ASP:187, GLU:166 |

**Figure 4**

***Docking visualisation of COVID-19 protease (6LU7) with anti-HIV drugs, Maraviroc (A), Etravirine (B), Indinavir (C) and Rilpivirine (D).***

**(A) Maraviroc**

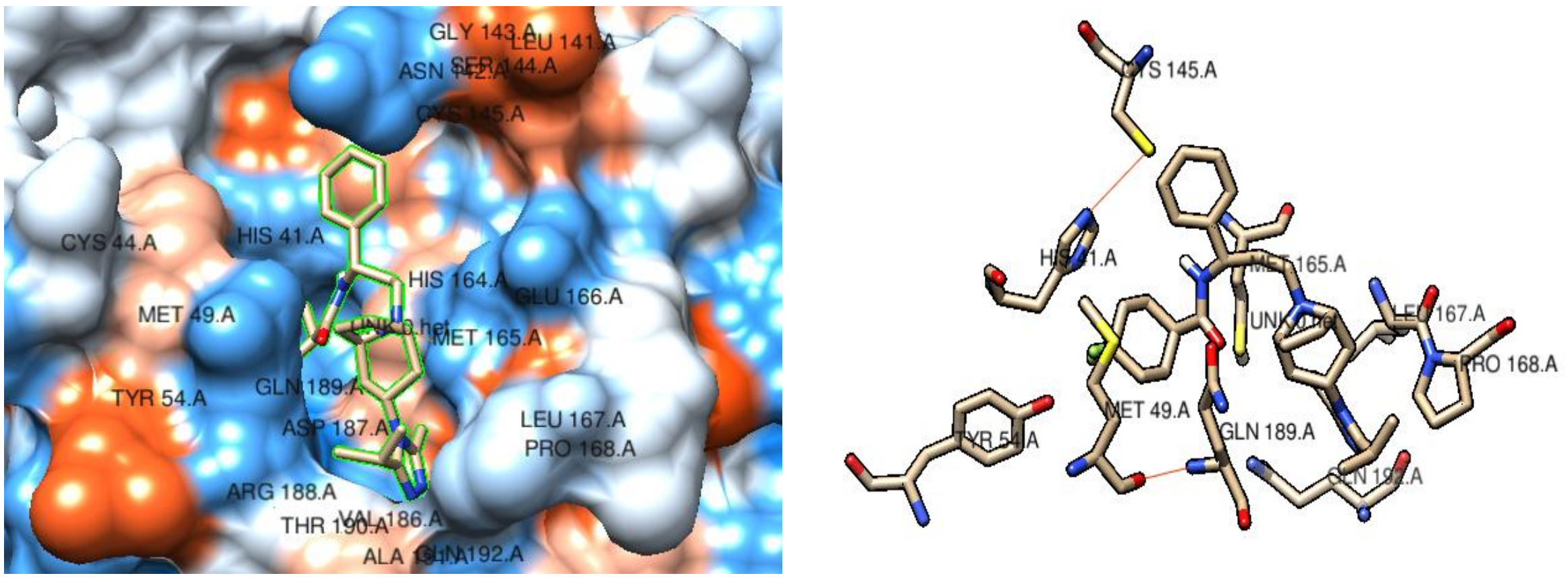

**(B) Etravirine**

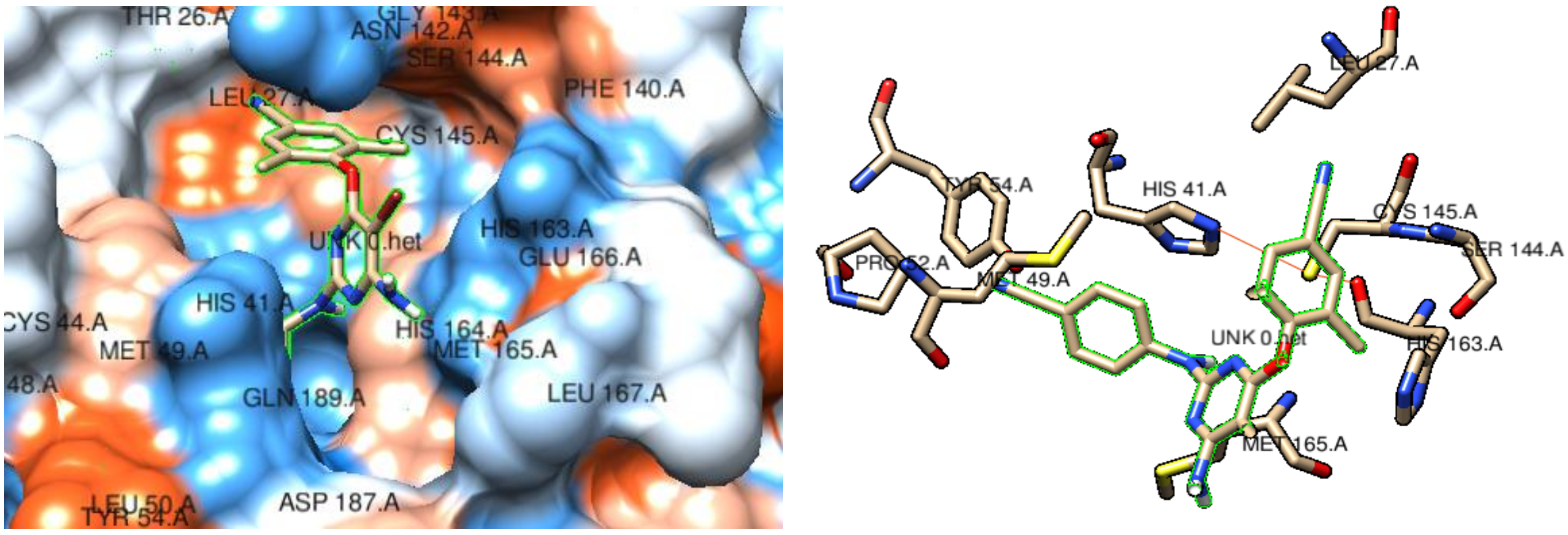

(C) **Indinavir**

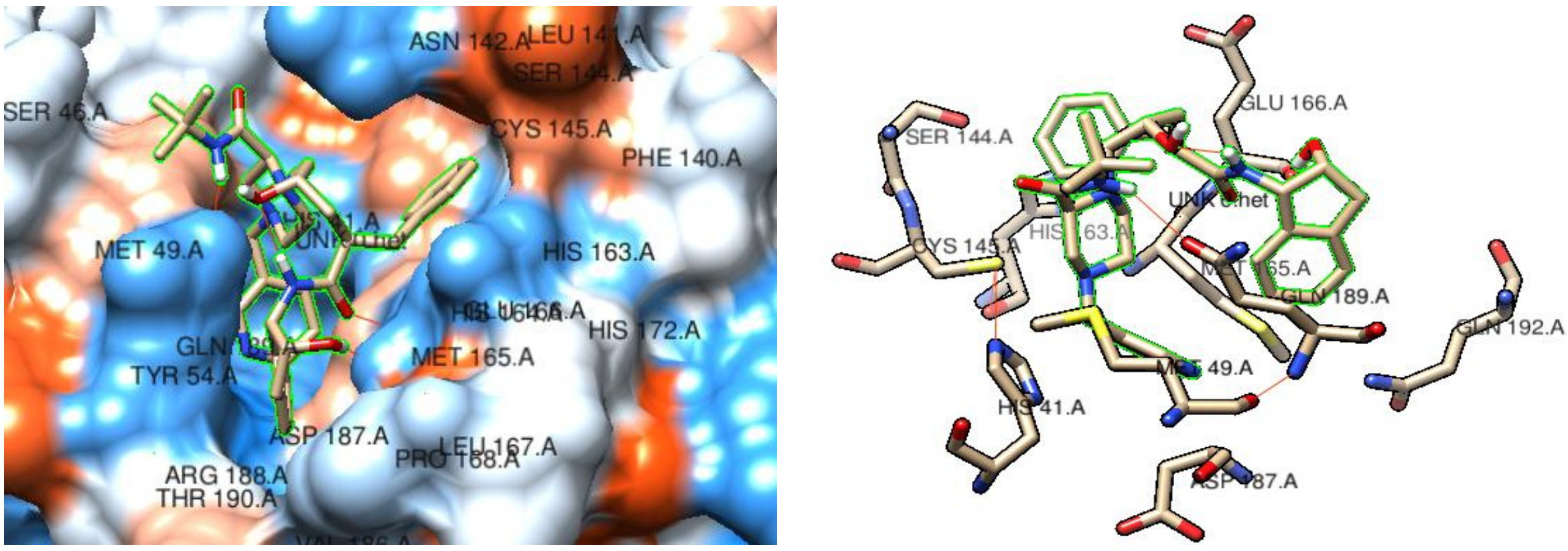

(D) **Rilpivirine**

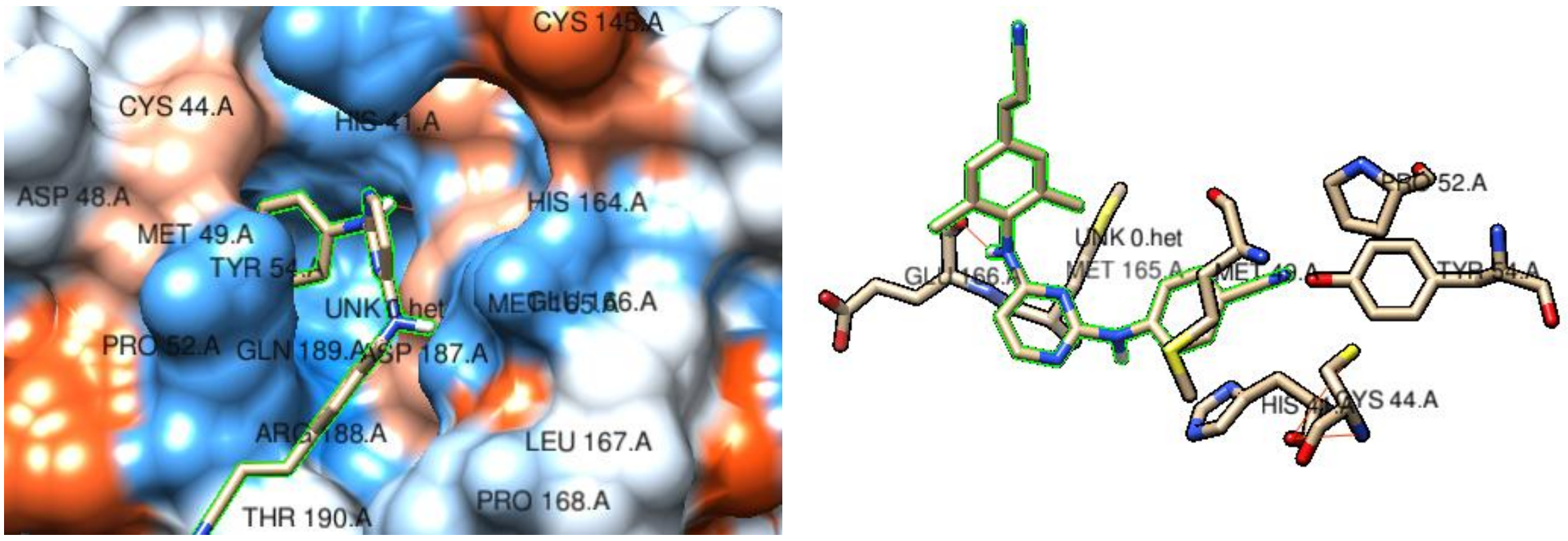

***Docking Prediction of anti-malarial drugs***

Seventeen clinically approved anti-malarial drugs were docked with SARS-CoV-2 protease. Out of seventeen, seven drugs were showed more potential inhibitors of SARS-CoV-2 main protease. Atovaquone, Quinidine, Halofantrine, Amodiaquine, Tetracycline, Azithromycin and hydroxychloroquine docked with binding affinity -8.95, -8.84, -8.68, -8.65, -8.4, -8.32 and -8.30 kcal/mol against SARS-CoV-2 main protease (Table 4). Docking visualization of 6LU7 with seven anti-malarial drugs was showed in Figure 5. Among the seven potent inhibitors Atovaquone was more potent with binding affinity -8.95 kcal/mol. Asn:142, Glu:166, Leu:141, Met:49, 165, Cys:44, Pro:52, Tyr:54, Arg:188, Asp:187, His:41,164, 163, 172, Phe:140 and Gln:189 are the amino acid residues involved in the interaction with Atovaquone (Figure 5). As these residues play an important role during protein-ligand interaction, they may serve as a biomarker during the drug discovery process.[18]

**Table 5**

***Molecular docking analysis of anti-malarial drugs against COVID-19 Protease (6LU7)***

| Sl. N | Compound Name | Binding energy (kcal/mol) | Residue involving interaction | No. of H bonds | Interaction of residues forming $H_2$ bonds |
|---|---|---|---|---|---|
| 1 | Quinine | -7.86 | GLN:189, HIS:172,41, 164, 163 CYS:145, MET:165, 49, GLU:166, ARG:188, TYR:54, ASN:142, PHE:140, SER:144, LEU:141, GLY:143, ASP:187 | 1 | GLN:189 |
| 2 | Quinidine | -8.84 | GLN:189, MET:165, 49, TYR:54, PRO:52, HIS:41, 164, PHE:181, ASP:187, GLU:166, ASN:142, CYS:145, SER:144, GLY:143, THR:26, 25, LEU:27, VAL:186 | 1 | GLN:189 |
| 3 | Mefloquine | -7.47 | GLN:189,192, TYR:54, GLU:166, ARG:188, ASP:187, MET:49, LEU:167, 141, THR:190, MET:165, HIS:163, 172, 164,41, LEU:167, CYS:145, PHE:140, SER:144, | 3 | GLN:189,192, TYR:54 |
| 4 | Chloroquine | -7.62 | HIS:164,163, 41, 172, ARG:188, PRO:52, TYR:54, MET:49, 165, ASP:187, GLN:189, GLU:166, LEU:141, SER:144, CYS:145, ASN:142, PHE:140 | 1 | HIS:164 |
| 5 | Amodiaquine | -8.65 | LEU:141,SER:144, HIS:41, 172, 163, 164, GLN:189, 192, MET:165,49,PHE:140, ASP:187, ARG:188, THR:190, GLN:192, GLU:166, ASN:142, GLY:143, CYS:145, | 2 | LEU:141, SER:144 |
| 6 | Primaquine | -7.15 | GLU:166, LEU:167, GLY:170, ASP:187, GLN:189, MET:165, ARG:188, GLN:192, THR:190, PRO:168 | 4 | GLU:166, LEU:167 |
| 7 | Halofantrine | -8.68 | GLN:192, 189, THR:190, ARG:188, GLU:166, CYS:145, MET:49, 165, HIS:164, ASP:187, TYR:54, PRO:168, LEU:167, 141,SER:144, HIS:163, GLY:143, | 2 | THR:190, GLN:192 |
| 8 | Sulfadoxine | -6.47 | HIS:164,41, 172, 163, SER:144, LEU:141, CYS:145, MET:49, 165 GLU:166, PRO:52, TYR:54, ARG:188, ASP:187, GLN:189, ASN:142 | 3 | SER:144, LEU:141, HIS:164 |
| 9 | Sulfamethoxypyridazine | -7.40 | HIS:163,164, 41,172, SER:144,LEU:141,MET:49, 165, ASP:187,ASN:142, PHE:140, GLU:166, GLN:189, TYR:54, PRO:52, CYS:145, ASN:142, GLY:143, | 4 | HIS:163,164, SER:144, LEU:141 |
| 10 | Proguanil | -7.81 | HIS:163, 172 GLU:166, LEU:141, PHE:140, MET:165, SER:144, GLY:143, ASN:142, CYS:145 | 4 | HIS:163 |
| 11 | Pyrimethamine | -6.85 | ASN:142, GLU:166, PHE:140, HIS:172, 163, 41, LEU:141, SER:144, MET:165, CYS:145, MET:49, GLY:143, | 3 | ASN:142, GLU:166, PHE:140 |
| 12 | Tetracycline | -8.40 | GLU:166, ASN:142, LEU:141, SER:144, CYS:145, MET:165,49, GLN:189, ASP:187, ARG:188, PHE:140, HIS:41,163,172, GLY:143 | 6 | GLU:166, LEU:141, SER:144, CYS:145, ASN:142, |
| 13 | Doxycycline | -8.30 | GLN:189, ASN:142, GLU:166, SER:144, MET:165, HIS:172, 163, PHE:140, LEU:141, CYS:145 | 6 | GLN:189, ASN:142, GLU:166, SER:144 |
| 14 | Azithromycin | -8.32 | ILE:152, PHE:294, 8, ARG:198, PRO:9, VAL:297, SER:301, ASP:153, TYR:154, ASN:151 | 1 | ILE:152 |
| 15 | Artemisinin | -7.63 | MET:165, 49, HIS:41, 164, ARG:188, GLU:166, GLN:189, CYS:145, 44, TYR:54, ASP:187, | 0 | 0 |
| 16 | Atovaquone | -8.95 | ASN:142, GLU:166, LEU:141, MET:49, 165, CYS:44, PRO:52, TYR:54, ARG:188, ASP:187, HIS:41,164, 163, 172, PHE:140, GLN:189 | 1 | ASN:142 |
| 17 | Hydroxychloroquin e | -8.30 | GLN:189, ARG:188, MET:165, TYR:54, ASP:187, HIS:41,172,163,164, PHE:140, LEU:141, CYS:145, SER:144, ASN:142, GLY:143, GLU:166 | 1 | GLN:189 |

**Figure 5.**

***Docking visualisation of COVID-19 protease (6LU7) with anti-malarial drugs, Atovaquone (A), Quinidine (B), Halofantrine (C), Amodiaquine (D), Tetracycline (E), Azithromycin (F) and hydroxychloroquine (G)***

**(A) Atovaquone**

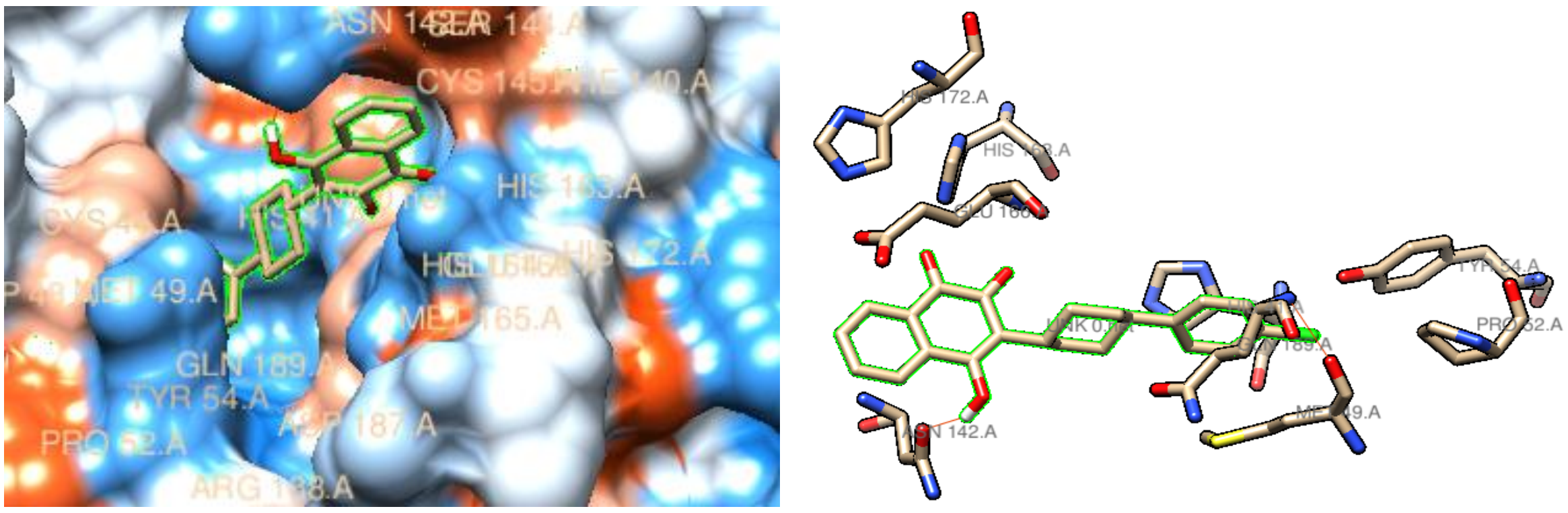

**(B) Quinidine**

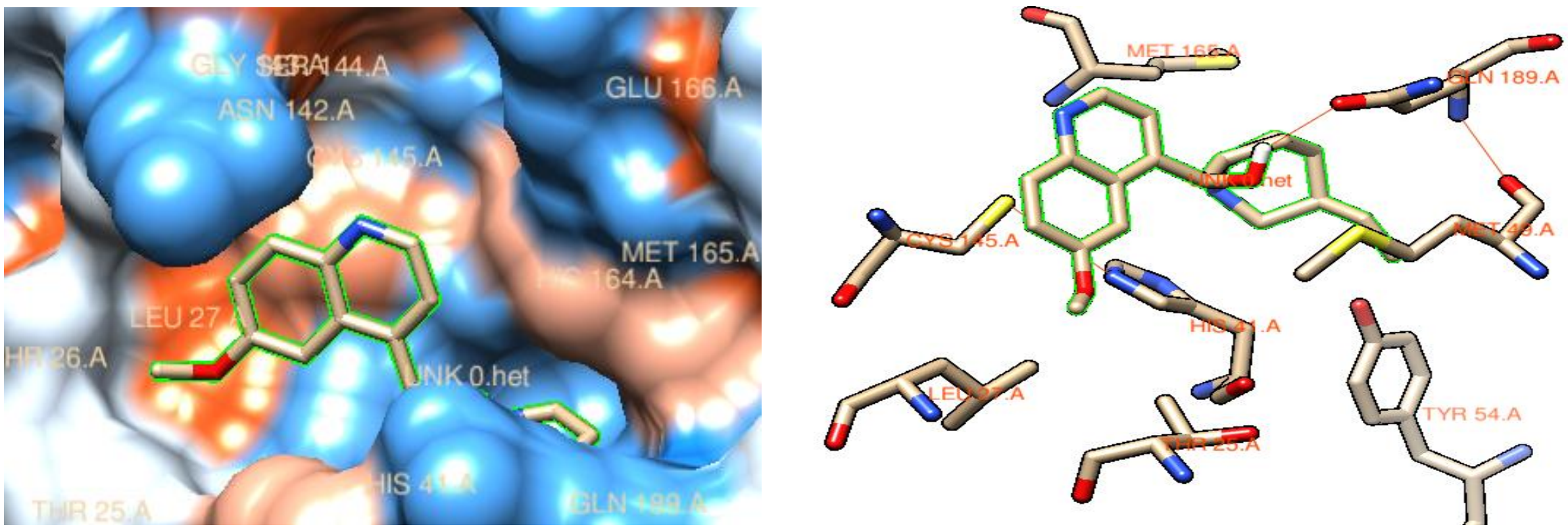

**(C) Halofantrine**

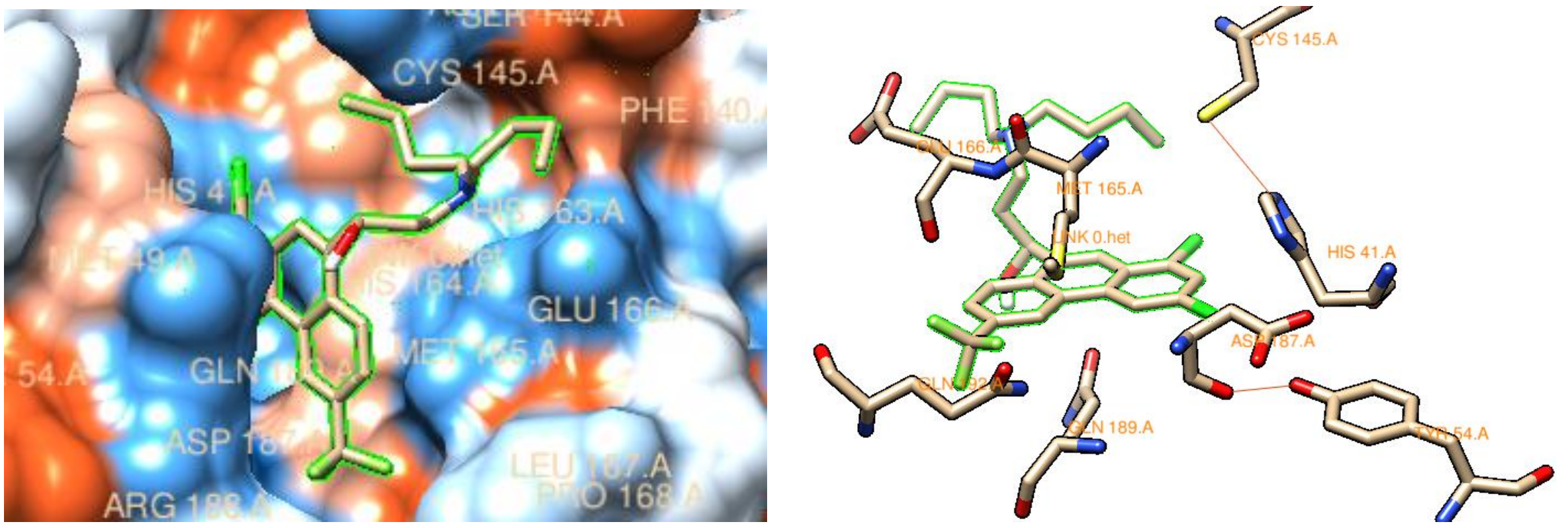

**(D) Amodiaquine**

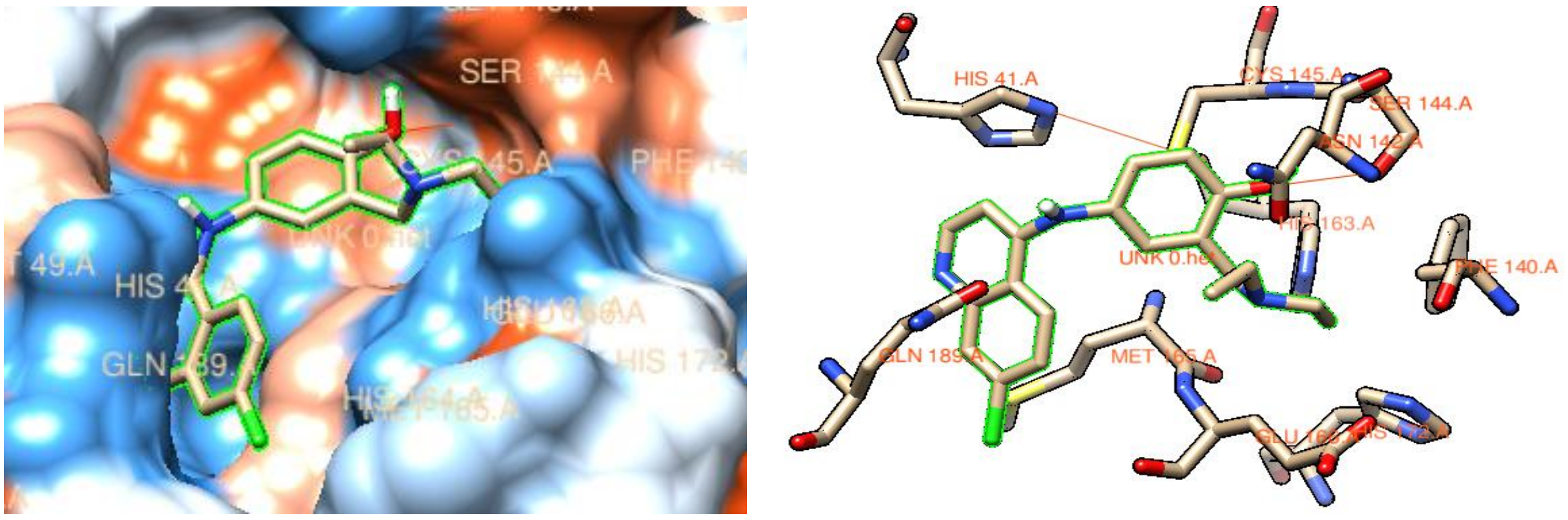

**(E) Tetracycline**

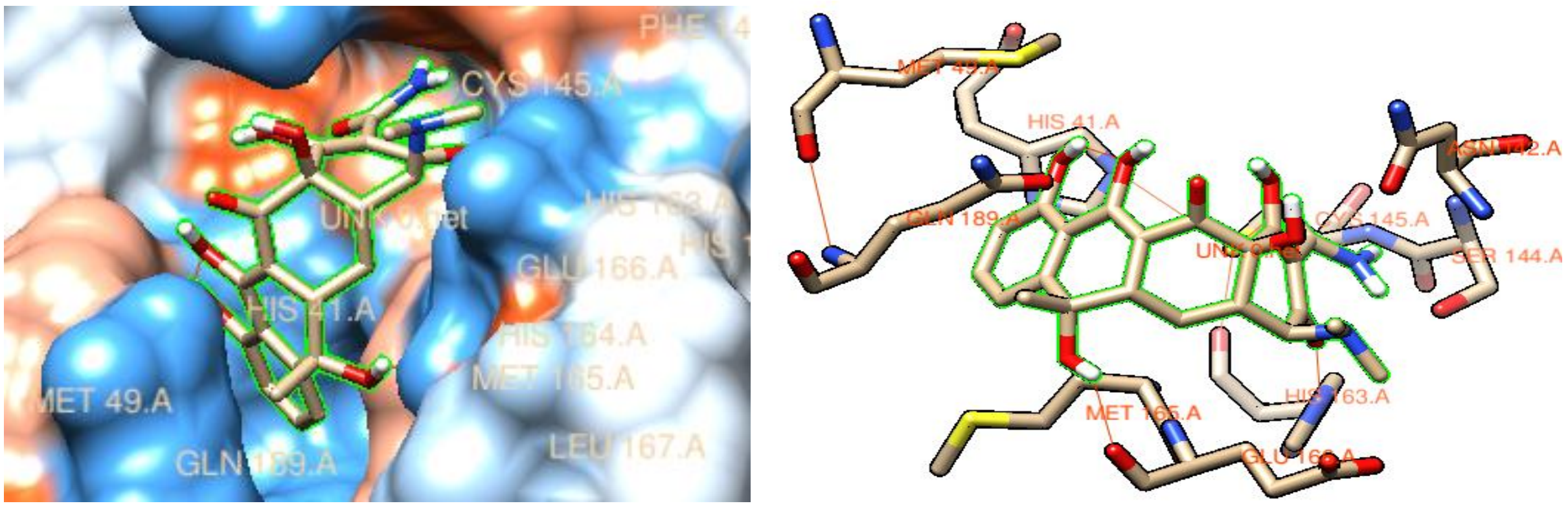

**(F) Azithromycin**

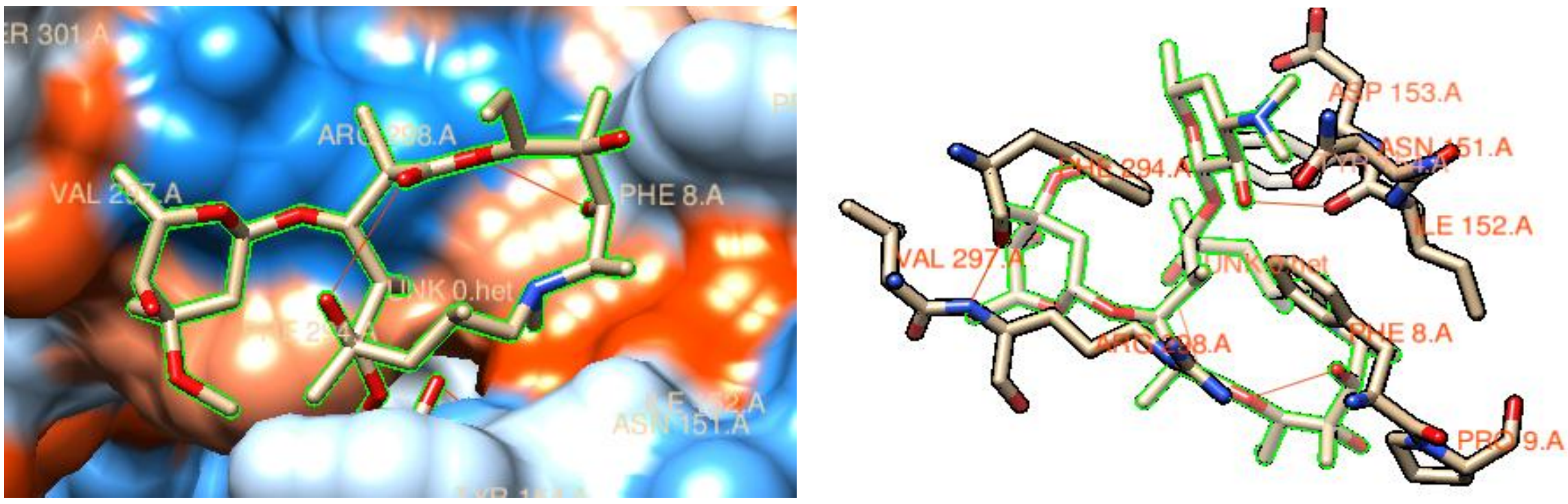

**(G) Hydroxychloroquine**

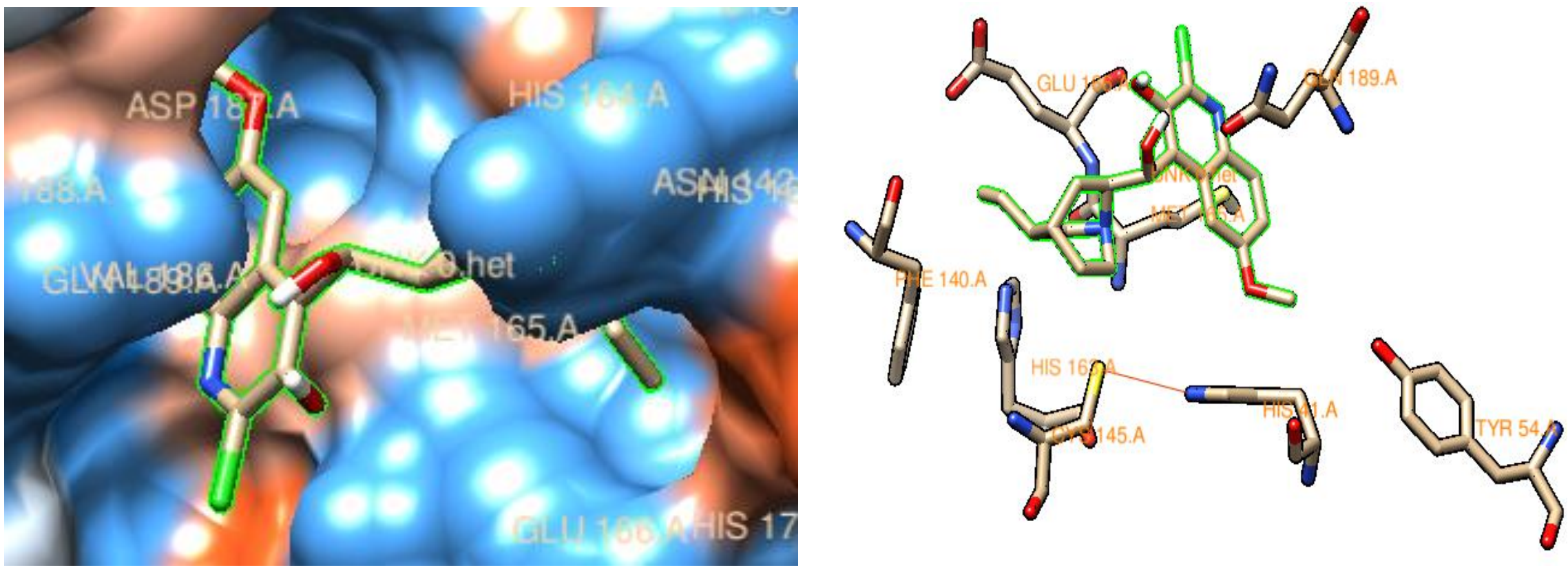

This study focused on identification of potential inhibitors against SARS-CoV-2 from corona virus to control the viral replication. Outcomes from the *In silico* molecular docking study maintained the great inhibitory efficacy of the one anti-H1N1 drug (Oseltamivir), one anti-TB drug (Rifampin), four anti-HIV drugs (Maraviroc, Etravirine, Indinavir, Rilpivirine) and seven anti-malarial drugs (Atovaquone, Quinidine, Halofantrine, Amodiaquine, Tetracylcine, Azithromycin, hydroxycholoroquine) since they could launch $H_2$ bonds with different amino acid residues that caused in an inhibition of SARS-CoV-2 protease activity with higher binding affinity ranging from (-10.67 to -8.3 kcal/mol). Thus, the projected binding interactions of the dynamic molecules with the protease by the docking study evidently established their inhibitory strength towards catalytic response of the protease. This study provides the support of the repurposed drugs, which may be helpful for the treatment of novel coronavirus disease and can serve as potential drug candidates to curb the ongoing and ever enlarging COVID-19 pandemic. Since all the drugs used in this study are of known pharmacokinetics standards and approved by FDA for human use they do not need to undergo specific long term clinical trials and therefore can fasten up the process of the therapeutics development.

## DISCUSSION

Coronavirus fits to a set of viruses which can contaminate vertebrate animals and human. It has slaughtered thousands of individuals around the world with growth in mortality rate each single day. Digestive, central nervous system, liver, respiratory systems of humans and animlas hampered by this virus infection.[19] Our study was focused on the FDA approved drugs against the main protease in coronavirus, as a possible beneficial target for the management of coronavirus. 6LU7 (PDB ID) is the major protease in COVID-19 that has been relocated and structured in PDB newly and is available to everybody in the world (Figure 1). For the proteolytic maturation of virus, the protease is precise significant. Protease has been studied as a possible target to avoid the extent of contamination by inhibiting viral polyprotein cleavage via blocking active sites of the protein. With this new finding of protease assembly in COVID-19, has providing an enormous chance to recognize possible drug candidates for the management of coronavirus.[20] In this study, we have applied a computational approach of FDA approved drugs in order to find a specific therapeutic possible agent against COVID-19. We have selected 47 FDA approved antiviral, anti-H1N1, anti-TB and anti-viral drugs and retrived directly from the PubChem (National Library of Medicine). Molecular docking was accomplished with the 47 drugs against COVID-19 structure.

Molecular docking is a computational technique which aims to find non-Covalent binding among protein (receptor) and a ligand/inhibitor (small molecule). For recognized binding site, the docking expects the method of interaction among a target protein and a ligand. Binding energy proposes the attraction of an exact ligand and asset by which a ligand interacts with and binds to the pocket of a target protein. A drug with a lesser binding energy (ΔG) is chosen as a probable drug candidate. In order to recognize the effect of active antiviral drugs on COVID-19, 47 FDA approved antiviral compounds were selected and performed molecular docking against COVID-19. Docking results of SARS-CoV-2 protease with selected 47 drugs out of the selected 13 showed best docking score and were found to be best molecules at the target site of the protein. Out of the 13 drugs, Maraviroc exhibited the best docked score (-10.67 kcal/mol) with SARS-CoV-2 protease. HIS:164, 41, TYR:54, ASP:187, MET:165, 49, LEU:167,141, PRO:168, CYS:44, 145, ARG:188, GLY:143, ASN:142, ALA:191, GLN:192, 189 and THR:190 are the amino acid residues participating in the interaction at the binding pocket of SARS-CoV-2 protease (Figure 4A).

Maraviroc is an effective anti-retroviral agent permitted for the treatment of HIV-1 infection that blocks interaction among the virus and the CCR5 co-receptor, a critical step in the HIV-1 replication. Earlier clinical trials of this drug have established the efficacy, tolerability, and safety of maraviroc in both treatment-naive and treatment-experienced patients.[21,22] Among the 17 anti-malarial approved drugs, Atovaquone presented best docking score (-8.95 kcal/mol) and ASN:142, GLU:166, LEU:141, MET:49, 165, CYS:44, PRO:52, TYR:54, ARG:188, ASP:187, HIS:41,164, 163, 172, PHE:140, GLN:189 are the amino acid residues participating in the interaction at the binding pocket of SARS-CoV-2 protease (Figure 5A). Recent studies also explores that, Atovaquone significantly inhibited ZIKV (Zika virus) in human placental JEG3 cells *in vitro*.[23]

## CONCLUSION

The present study concludes that, thirteen clinically approved drugs are identified as potent inhibitors against SARS-CoV-2 protease activity. These outcome afford a strong foundation for the use of these drugs are for CORONA management. Moreover, the dynamic ligands inhibited the catalytic response of protease by blocking the residues of amino acids intricate in the processing and strand transmission reactions. The interactions by the structural model at the protease active site can afford a valuable guide for additional strategies for structure-based medicines and development of new operative inhibitors of SARS-CoV-2 protease. Therefore, the effect of these inhibitors can be further revealed through *in vitro* and *in vivo* analysis in the termination of intracellular replication of corona virus, prior to the use as drugs in humans.

## ACKNOWLEDGEMENTS

All the authors are gratefully acknowledge to Department of Zoology, Kakatiya University, Warangal, Telangana, India for providing Bioinformatics lab to carry out this study. No funding to declare.

## CONFLICT OF INTEREST

Conflict of interest declared none.